\documentclass[conference,letterpaper]{IEEEtran}

\usepackage{cite}
\usepackage{latexsym}
\usepackage{amsmath, amssymb}
\usepackage{mathrsfs}
\usepackage[dvips,dvipdf]{graphicx}
\usepackage{subfigure}
\usepackage{algorithm}
\usepackage{algorithmic}

\begin{document}

\newcommand{\A}{\mathbf{A}}
\newcommand{\B}{\mathbf{B}}
\newcommand{\C}{\mathbf{C}}
\newcommand{\D}{\mathbf{D}}
\renewcommand{\d}{\mathbf{d}}
\newcommand{\e}{\mathbf{e}}
\newcommand{\E}{\mathbf{E}}
\newcommand{\f}{\mathbf{f}}
\newcommand{\F}{\mathbf{F}}
\newcommand{\G}{\mathbf{G}}
\newcommand{\GG}{\mathbf{\Gamma}}
\newcommand{\I}{\mathbf{I}}
\newcommand{\M}{\mathbf{M}}
\newcommand{\N}{\mathbf{N}}
\newcommand{\n}{\mathbf{n}}
\newcommand{\Q}{\mathbf{Q}}
\newcommand{\q}{\mathbf{q}}
\newcommand{\R}{\mathbf{R}}
\renewcommand{\r}{\mathbf{r}}
\renewcommand{\S}{\mathbf{S}}
\newcommand{\T}{\mathbf{T}}
\newcommand{\s}{\mathbf{s}}
\renewcommand{\t}{\mathbf{t}}
\newcommand{\U}{\mathbf{U}}
\renewcommand{\u}{\mathbf{u}}
\renewcommand{\v}{\mathbf{v}}
\newcommand{\W}{\mathbf{W}}
\newcommand{\X}{\mathbf{X}}
\newcommand{\x}{\mathbf{x}}
\newcommand{\Y}{\mathbf{Y}}
\newcommand{\y}{\mathbf{y}}
\newcommand{\z}{\mathbf{z}}
\newcommand{\rb}{\mathbf{b}}
\newcommand{\rvec}{\mathbf{r}}
\newcommand{\0}{\mathbf{0}}
\newcommand{\1}{\mathbf{1}}
\newcommand{\Hc}{\mathbf{H}}
\newcommand{\DL}{\mathbf{DL}}
\newcommand{\mLambda}{\mathbf{\Lambda}}
\newcommand{\mOmega}{\mathbf{\Pi}}
\newcommand{\valpha}{\mathbf{\alpha}}
\newcommand{\Sone}{\mathcal{S}}
\newcommand{\tr}{\mathrm{Tr}}
\newcommand{\new}{\mathrm{new}}
\newcommand{\net}{\mathrm{net}}
\newcommand{\link}{\mathrm{link}}
\newcommand{\LL}{\mathbf{L}}
\newcommand{\Out}[1]{\mathcal{O}\left( #1 \right)}
\newcommand{\In}[1]{\mathcal{I}\left( #1 \right)}
\newcommand{\Int}[1]{\mathcal{F}\left( #1 \right)}
\newcommand{\sInt}[1]{\hat{\mathcal{F}}\left( #1 \right)}
\renewcommand{\Re}[1]{\mbox{$\mathfrak{Re} \left( #1 \right)$}}
\renewcommand{\Im}[1]{\mbox{$\mathfrak{Im} \left( #1 \right)$}}
\newcommand{\Resq}[1]{\mbox{$\mathfrak{Re}^{2} \left( #1 \right)$}}
\newcommand{\Imsq}[1]{\mbox{$\mathfrak{Im}^{2} \left( #1 \right)$}}
\newcommand{\LTM}{\langle L \rangle}
\newcommand{\con}[1]{\overline{#1}}
\newcommand{\abs}[1]{\mbox{$\lvert #1 \rvert$}}
\newcommand{\norm}[1]{\mbox{$\left\lVert #1 \right\rVert$}}
\newcommand{\dual}{$\mathbf{D}^{\mathrm{CRPA}}$}
\newcommand{\dnet}{$\mathbf{D}^{\mathrm{CRPA}}_{\net}$}
\newcommand{\dlink}{$\mathbf{D}^{\mathrm{CRPA}}_{\link}$}
\newcommand{\Pwrcon}{\Omega_{+}(P_{\max}^{(n)})}
\newcommand{\src}[1]{\mathrm{src}(#1)}
\newcommand{\dst}[1]{\mathrm{dst}(#1)}
\newcommand{\rx}[1]{Rx(#1)}
\newcommand{\tx}[1]{Tx(#1)}
\newcommand{\mtrx}[1]{\mbox{$\left[\begin{array} #1 \end{array}\right]$}}
\newcommand{\diag}[1]{\mbox{Diag}\mbox{$\left\{ #1 \right\}$}}
\newcommand{\Cmac}{\mathcal{C}_{\mathrm{MAC}}}
\newcommand{\Cbc}{\mathcal{C}_{\mathrm{DPC}}}
\newcommand{\ddet}[1]{\left| #1 \right|}

\newtheorem{thm}{Theorem}
\newtheorem{cor}[thm]{Corolary}
\newtheorem{lem}{Lemma}
\newtheorem{prop}[thm]{Proposition}
\newtheorem{ex}{Example}[section]
\newtheorem{defn}[thm]{Definition}
\newtheorem{finalremark}[thm]{Final Remark}
\newtheorem{rem}{Remark}
\newtheorem{sol}{Solution}



\title{Cross-Layer Optimization of MIMO-Based Mesh Networks with Gaussian Vector Broadcast Channels}

\author{
\authorblockN{Jia Liu and Y. Thomas Hou}
\authorblockA{The Bradley Department of Electrical and Computer Engineering\\
Virginia Polytechnic Institute and State University, Blacksburg, VA 24061
\\Email: \{kevinlau, thou\}@vt.edu}
 }

\maketitle
\begin{abstract}
MIMO technology is one of the most significant advances in the past decade to increase channel
capacity and has a great potential to improve network capacity for mesh networks. In a MIMO-based
mesh network, the links outgoing from each node sharing the common communication spectrum can be
modeled as a Gaussian vector broadcast channel. Recently, researchers showed that ``dirty paper
coding'' (DPC) is the optimal transmission strategy for Gaussian vector broadcast channels. So far,
there has been little study on how this fundamental result will impact the cross-layer design for
MIMO-based mesh networks. To fill this gap, we consider the problem of jointly optimizing DPC power
allocation in the link layer at each node and multihop/multipath routing in a MIMO-based mesh
networks. It turns out that this optimization problem is a very challenging non-convex problem. To
address this difficulty, we transform the original problem to an equivalent problem by exploiting
the channel duality. For the transformed problem, we develop an efficient solution procedure that
integrates Lagrangian dual decomposition method, conjugate gradient projection method based on
matrix differential calculus, cutting-plane method, and subgradient method. In our numerical
example, it is shown that we can achieve a network performance gain of $34.4\%$  by using DPC.
\end{abstract}


\section{Introduction}
Since Telatar's \cite{Telatar99:Limits} and Foschini's \cite{Foschini98:Limits} pioneering works
predicting the potential of high spectral efficiency provided by multiple antenna systems, the last
decade has witnessed a soar of research activity on Multiple-Input Multiple-Output (MIMO)
technologies. The benefits of substantial improvements in wireless link capacity at no cost of
additional spectrum and power have quickly positioned MIMO as one of the breakthrough technologies
in modern wireless communications, rendering it as an enabling technology for next generation
wireless networks. However, applying MIMO in wireless mesh networks (WMNs) is not a trivial
technical extension. With the increased number of antennas at each node, interference is likely to
become stronger if power level at each node, power allocation to each antenna element, and routing
are not managed wisely. As a result, cross-layer design is necessary for MIMO-based WMNs.

\begin{figure}[ht!]
\centerline{\subfigure[Time or frequency division]{\includegraphics[width=1.6in]{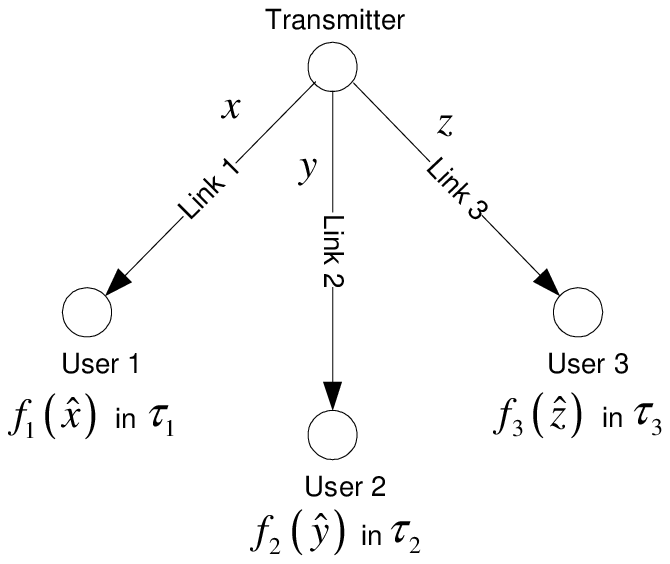}
\label{fig_TDM}} \hspace{-.1in} \hfil \subfigure[DPC transmission
strategy]{\includegraphics[width=1.6in]{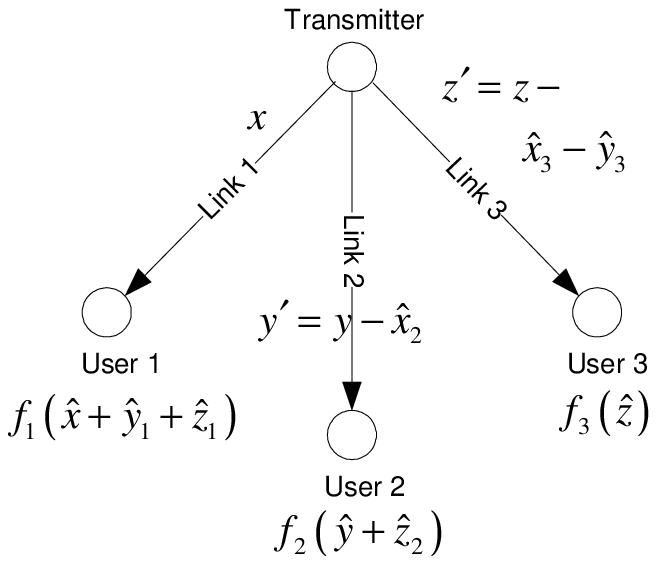} \label{fig_DPC}} } \caption{A 3-user broadcast
channel example.} \label{fig_bc_example}
\end{figure}

In a MIMO-based WMN, the set of outgoing links from a node sharing a common communication spectrum
can be modeled as a nondegraded Gaussian vector broadcast channel, for which the capacity region is
notoriously hard to analyze \cite{Cover91:Info_Thry}. In the networking literature, most works
considering links sharing a common communication spectrum are concerned with how to allocate
frequency sub-bands/time-slots and schedule transmissions to efficiently share the common
communication spectrum. As an example, Fig.~\ref{fig_TDM} shows a simple broadcast channel where
there are three uncoordinated users and a single transmitting node. Suppose that messages $x$, $y$,
and $z$ need to be delivered to user 1, user 2, and user 3, respectively. Also, suppose that the
received signals subject to ambient noise are $\hat{x}$, $\hat{y}$, and $\hat{z}$, and the decoding
functions are $f_{1}(\cdot)$, $f_{2}(\cdot)$, and $f_{3}(\cdot)$, respectively. The conventional
strategy is to divide a unit time frame into three time slots (or divide a unit band into three
sub-bands) $\tau_{1}$, $\tau_{2}$, and $\tau_{3}$, and then find the optimal scheduling for
transmissions to users 1, 2 and 3, accordingly. The major benefit of this strategy is that
interference can be eliminated.

Although the time or frequency division schemes are simple and effective, they are not necessarily
the smartest strategy. In fact, Cover had shown in his classical paper \cite{Cover72:BC} that the
transmission scheme {\em jointly} encoding all receivers' information at the transmitter can do
strictly better in broadcast channels. However, the capacity achieving transmission signaling
scheme for general nondegraded Gaussian vector broadcast channels is very difficult to determine
and has become one of the most basic questions in network information theory
\cite{Cover91:Info_Thry}. Very recently, significant progress has been made in this area. Most
notably, Weigarten {\em et. al.} finally proved the long-open conjecture that the ``dirty paper
coding'' strategy (DPC) \cite{Costa03:DPC} is the optimal transmission scheme for Gaussian vector
broadcast channels \cite{Weingarten06:MIMO_BC} in the sense that the DPC rate region $\Cbc$ of a
broadcast channel is equal to the broadcast channel's capacity region $\mathcal{C}_{\mathrm{BC}}$,
i.e., $\mathcal{C}_{\mathrm{BC}} = \Cbc$. However, this fundamental result is still not adequately
exposed to the networking research community. So far, how to exploit DPC's benefits in the
cross-layer design for wireless mesh networks has not yet been studied in the literature. The main
objective of this study is to fill this gap and to obtain a rigorous and systematic understanding
of the impact of applying DPC to the cross-layer optimization for MIMO-based mesh networks.

To begin with, it is beneficial to introduce the basic idea of DPC, which turns out to be very
simple. For the same 3-user example, consider the following strategy as shown in
Fig.~\ref{fig_DPC}. We first jointly encode the messages for all the users in a certain order and
then broadcast the resulting codeword {\em simultaneously}. Suppose that we pick user 1 to be
encoded first, then followed by user 2, and finally user 3. We choose the codeword $x$ for user 1
as before. Then, the interference seen by user 2 due to user 1 (denoted by $\hat{x}_{2}$) is known
at the transmitter. So, the transmitter can subtract the interference and encode user 2 as $y' = y
- \hat{x}_{2}$ rather than $y$ itself. As a result, user 2 does not see any interference from the
signal intended for user 1. Likewise, after encoding user 2, the interferences seen by user 3 due
to user 1 and 2 (denoted by $\hat{x}_{3}$ and $\hat{y}_{3}$) are known at the transmitter. Then,
the transmitter can subtract the interferences and encode user 3 as $z' = z - \hat{x}_{3} -
\hat{y}_{3}$ rather than $z$ itself. Therefore, user 3 does not see any interferences from the
signals intended for user 1 and 2. In the end, the transmitter adds all the codewords together and
broadcasts the sum to all users simultaneously. As a result, it is easy to see from
Fig.~\ref{fig_DPC} that the received signal at user 1 is $\hat{x} + \hat{y}_{1} + \hat{z}_{1}$,
i.e., user 1 will experience the interference from the signals intended for users 2 and 3; the
received signal at user 2 is $\hat{y} + \hat{z}_{2}$, i.e., user 2 only experiences the
interference from the signal intended for user 3; and finally, the received signal at user 3 is
$\hat{z}$, i.e., user 3 does {\em not} experience any interference. This process operates like
writing on a dirty paper, hence the name. Although counterintuitive, the capacity region of DPC
that allows interference is strictly larger than those of time or frequency division schemes.

After understanding what DPC is, one may ask two very natural and interesting questions:
\begin{enumerate}
\item {\em How will the enlarged capacity region at each node due to DPC impact the network
performance in the upper layers?}

\item {\em Are there any new challenges if DPC is employed in a MIMO-based networking environment?}
\end{enumerate}

Notice that, when DPC is employed, the encoding order plays a critical role. For a $K$-user
broadcast channel, there exists $K!$ permutations. Also, since DPC allows interference among the
users, power allocation among different users along with the encoding order has a significant
impact on the system performance. As we show later, the DPC link rates in a broadcast channel are
non-connvex functions. Thus, even the optimization for a single $K$-user Gaussian vector broadcast
channel is a very challenging {\em combinatorial non-convex} problem, not to mention the
cross-layer design in a networking environment with multiple broadcast channels.

In this paper, we aim to solve the problem of jointly optimizing DPC per-antenna power allocation
at each node in the link layer and multihop/multipath routing in a MIMO-based WMN. Our
contributions are three-fold. First, this paper is the first work that studies the impacts of
applying DPC to the cross-layer design for MIMO-based WMNs. In our numerical example, it is shown
that we can achieve a network performance gain of $34.4\%$ by using DPC in MIMO-based WMNs. Also,
since the traditional single-antenna systems can be viewed as a special case of MIMO systems, the
findings and results in this paper are also applicable to conventional WMNs with single-antenna.
Second, to address the non-convex difficulty, we transform the original problem to an equivalent
problem under the dual MIMO multiple access channel (MIMO-MAC) and show that the transformed
problem is convex with respect to the input covariance matrices. We simplify the maximum weighted
sum rate problem for the dual MIMO-MAC such that enumerating different encoding order is
unnecessary, thus paving the way to efficiently solve the link layer subproblem in Lagrangian dual
decomposition. Last, for the transformed problem, we develop an efficient solution procedure that
integrates Lagrangian dual decomposition method, conjugate gradient projection method based on
matrix differential calculus, cutting-plane method, and subgradient method.

The remainder of this paper is organized as follows. In Section~\ref{sec:model_formulation}, we
discuss the network model and problem formulation. Section~\ref{sec:reformulation} discusses how to
reformulate the non-connvex original problem by exploiting channel duality. In
Section~\ref{sec:soln_proc}, we introduce the key components for solving the challenging link layer
subproblem in the Lagrangian decomposition. Numerical results are provided in
Section~\ref{sec:numerical} to illustrate the efficacy of our proposed solution procedure and to
study the network performance gain by using DPC. Section~\ref{sec:related} reviews related work and
Section~\ref{sec:conclusion} concludes this paper.

\section{Network Model} \label{sec:model_formulation}
We first introduce notations for matrices, vectors, and complex scalars in this paper. We use
boldface to denote matrices and vectors. For a matrix $\A$, $\A^{\dag}$ denotes the conjugate
transpose, $\tr\{\A\}$ denotes the trace of $\A$, and $|\A|$ denotes the determinant of $\A$.
$\diag{\A_{1}, \ldots,\A_{n}}$ represents the block diagonal matrix with matrices
$\A_{1},\ldots,\A_{n}$ on its main diagonal. We let $\I$ denote the identity matrix with dimension
determined from context. $\A \succeq 0$ represents that $\A$ is Hermitian and positive semidefinite
(PSD). $\1$ and $\0$ denote vectors whose elements are all ones and zeros, respectively, and their
dimensions are determined from context. $(\v)_{m}$ represents the $m^{th}$ entry of vector $\v$.
For a real vector $\v$ and a real matrix $\A$, $\v \geq \0$ and $\A \geq \0$ mean that all entries
in $\v$ and $\A$ are nonnegative, respectively. We let $\e_{i} $ be the unit column vector where
the $i^{th}$ entry is $1$ and all other entries are $0$. The dimension of $\e_{i}$ is determined
from context as well. The operator ``$\langle , \rangle$'' represents the inner product operation
for vectors or a matrices.

\subsection{Network Layer} \label{sec:network_flow}
In this paper, the topology of a MIMO-based wireless mesh network is represented by a directed
graph, denoted by $\mathcal{G} = \{ \mathcal{N}, \mathcal{L} \}$, where $\mathcal{N}$ and
$\mathcal{L}$ are the set of nodes and all possible MIMO-based links, respectively. By saying
``possible'' we mean the distance between a pair of nodes is less than or equal to the maximum
transmission range $D_{\max}$, i.e., $\mathcal{L} = \{ (i,j): D_{ij} \leq D_{\max}, \, i,j \in
\mathcal{N}, i \ne j\}$, where $D_{ij}$ represents the distance between node $i$ and node $j$.
$D_{\max}$ can be determined by a node's maximum transmission power. We assume that $\mathcal{G}$
is always connected. Suppose that the cardinalities of the sets $\mathcal{N}$ and $\mathcal{L}$ are
$|\mathcal{N}|=N$ and $|\mathcal{L}|=L$, respectively. For convenience, we index the links
numerically (e.g., link $1,2,\ldots,L$) rather than using node pairs $(i,j)$.

The network topology of $\mathcal{G}$ can be represented by a \emph{node-arc incidence matrix}
(NAIM) \cite{Bazaraa_Jarvis_Sherali_90:LP} $\A \in \mathbb{R}^{N \times L}$, whose entry $a_{nl}$
associating with node $n$ and arc $l$ is defined as
\begin{equation} \label{eqn_naim_a}
a_{nl} = \left\{ \begin{array}{rl} 1 & \mathrm{if} \,\, n \,\, \mathrm{is \,\, the \,\, transmitting \,\, node \,\, of \,\, arc \,\,} l \\
-1 & \mathrm{if} \,\, n \,\, \mathrm{is \,\, the \,\, receiving \,\, node \,\, of \,\, arc \,\,} l  \\
0 & \mathrm{otherwise.} \\
\end{array}\right.
\end{equation}
We define $\Out{n}$ and $\In{n}$ as the sets of links that are outgoing from and incoming to node
$n$, respectively. We use a multicommodity flow model for the routing of data packets across the
network. In this model, several nodes send different data to their corresponding destinations,
possibly through {\em multipath} and {\em multihop} routing. We assume that the flow conservation
law at each node is satisfied, i.e., the network is a flow-balanced system.

Suppose that there are $F$ sessions in total in the network, representing $F$ different
commodities. The source and destination nodes of session $f$, $1 \leq f \leq F$, are denoted as
$\src{f}$ and $\dst{f}$, respectively. For the supply and demand of each session, we define a
\emph{source-sink vector} $\s_{f} \in \mathbb{R}^{N}$, whose entries, other than at the positions
of $\src{f}$ and $\dst{f}$, are all zeros. In addition, from the flow conservation law, we must
have $(\s_{f})_{\src{f}} = -(\s_{f})_{\dst{f}}$. Without loss of generality, we let
$(\s_{f})_{\src{f}} \geq 0$ and simply denote it as a scalar $s_{f}$. Therefore, we can further
write the source-sink vector of flow $f$ as
\begin{equation} \label{eqn_src_sink_vec}
\s_{f} = s_{f} \left[ \begin{array}{ccccc} \cdots & 1 & \cdots & -1 & \cdots
\end{array} \right]^{T},
\end{equation}
where the dots represent zeros, and $1$ and $-1$ are in the positions of $\src{f}$ and $\dst{f}$,
respectively. Note that for the source-sink vector of a session $f$, $1$ does not necessarily
appear before $-1$ as in (\ref{eqn_src_sink_vec}), which is only for an illustrative purpose. Using
the notation ``$=_{x,y}$'' to represent the component-wise equality of a vector except at the
$x^{th}$ and the $y^{th}$ entries, we have $\s_{f} =_{\src{f},\dst{f}} \0$. In addition, using the
matrix $\S \triangleq \mtrx{{cccc} \s_{1} & \s_{2} & \ldots & \s_{F} } \in \mathbb{R}^{N \times F}$
to denote the collection of all source-sink vectors, we further have
\begin{eqnarray}
& \S \e_{f} =_{\src{f},\dst{f}} \0, & 1 \leq f \leq F, \\
& \langle \1, \S\e_{f} \rangle = 0, & 1 \leq f \leq F, \\
& (\S \e_{f})_{\src{f}} = s_{f}, & 1 \leq f \leq F,
\end{eqnarray}
where $\e_{f}$ is the $f^{th}$ unit column vector.

On link $l$, we let $t_{l}^{(f)} \geq 0$ be the amount of flow of session $f$ in link $l$. We
define $\t^{(f)} \in \mathbb{R}^{L}$ as the $\emph{flow vector}$ for session $f$. At node $n$,
components of the flow vector and source-sink vector for the same commodity satisfy the flow
conservation law as follows: $\sum_{l \in \Out{n}} t_{l}^{(f)} - \sum_{l \in \In{n}} t_{l}^{(f)} =
(\s_{f})_{n}$, $1 \leq n \leq N$, $1 \leq f \leq F$. With NAIM, the flow conservation law across
the whole network can be compactly written as $\A \t^{(f)} = \s_{f}, \quad 1 \leq f \leq F$. We use
matrix $\T \triangleq \mtrx{{cccc} \t^{(1)} & \t^{(2)} & \ldots & \t^{(F)} } \in \mathbb{R}^{L
\times F}$ to denote the collection of all flow vectors. With $\T$ and $\S$, the flow conservation
law can be further compactly written as $\A \T = \S$.

\subsection{Channel Capacity of a MIMO Link}
In this section, we first briefly introduce some background of MIMO. We use a matrix $\Hc_{l} \in
\mathbb{C}^{n_{r} \times n_{t}}$ to represent the MIMO channel gain matrix from the transmitting
node to the receiving node of link $l$, where $n_{t}$ and $n_{r}$ are the numbers of transmitting
and receiving antenna elements of each node, respectively. $\Hc_{l}$ captures the effect of the
scattering environment between the transmitter and the receiver of link $l$. In an additive white
Gaussian noise (AWGN) channel, the received complex base-band signal vector for a MIMO link $l$
with $n_{t}$ transmitting antennas and $n_{r}$ receiving antennas is given by
\begin{equation} \label{eqn_baseband}
\y_{l} = \sqrt{\rho_{l}} \Hc_{l} \x_{l} + \n_{l}.
\end{equation}
where $\y_{l}$ and $\x_{l}$ represent the received and transmitted signal vector; $\n_{l}$ is the
normalized additive white Gaussian noise vector; $\rho_{l}$ captures the path-loss effect, which is
usually modeled as $\rho_{l} = G \cdot D_{l}^{-\alpha}$, where $G$ is some system specific
constant, $D_{l}$ denotes the distance between the transmitting node and the receiving node of link
$l$, and $\alpha$ denotes the path loss exponent. Let matrix $\Q_{l}$ represent the covariance
matrix of a zero-mean Gaussian input symbol vector $\x_{l}$ at link $l$, i.e., $\Q_{l} = \mathbb{E}
\left\{ \x_{l} \cdot \x_{l}^{\dag} \right\}$. This implies that $\Q_{l}$ is Hermitian and $\Q_{l}
\succeq 0$. Physically, $\Q_{l}$ represents the power allocation in different antenna elements in
link $l$'s transmitter and the correlation between each pair of the transmit and receive antenna
elements. $\tr\{\Q_{l}\}$ is the total transmission power at the transmitter of link $l$. The
capacity of a MIMO link $l$ in an AWGN channel with a unit bandwidth can be computed as
\begin{eqnarray} \label{eqn_link_cap}
R_{l}(\Q_{l}) = \log_{2} \ddet { \I + \rho_{l} \Hc_{l} \Q_{l} \Hc_{l}^{\dag} },
\end{eqnarray}
It can be seen that different power allocations to the antennas will have different impacts on the
link capacity. Therefore, the optimal input covariance matrix $\Q_{l}^{*}$ needs to be determined.
In a single link environment, the optimal input covariance matrix can be computed by water-filling
the total power over the eigenmodes (signaling direction) of the MIMO channel matrix
\cite{Telatar99:Limits}. However, in a networking environment, finding the optimal input covariance
matrices is a substantially more challenging task. Determining the optimal input covariance
matrices is one of the major goals in our cross-layer optimization.

\subsection{MIMO-BC Link Layer} \label{sec:mimo_linkcap}
\begin{figure}[ht!]
\centering
\includegraphics[width=3.1in]{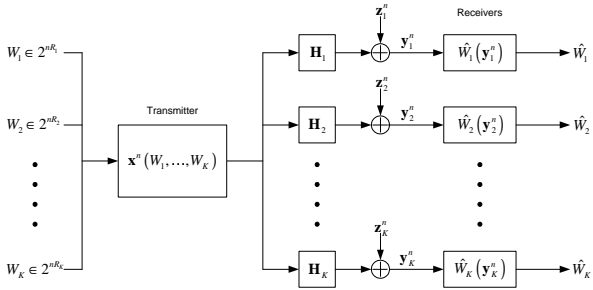}
\caption{A Gaussian vector broadcast channel.} \label{fig_bc}
\end{figure}

A communication system where a single transmitter sends independent information to multiple {\em
uncoordinated} receivers is referred to as a broadcast channel. If the channel gain of each link in
the broadcast channel is a matrix and the noise to each link is a Gaussian random vector, the
channel is termed ``Gaussian vector broadcast channel''. Fig.~\ref{fig_bc} illustrates a $K$-user
Gaussian vector broadcast channel, where independent messages $W_{1},\ldots, W_{K}$ are jointly
encoded by the transmitter, and the receivers are trying to decode $W_{1}, \ldots, W_{K}$,
respectively. A $(n, 2^{nR_{1}}, \ldots, 2^{nR_{K}})_{BC}$ codebook for a broadcast channel
consists of an encoding function $\x^{n}(W_{1}, \ldots, W_{K})$ where $W_{i} \in \{ 1,\ldots,
2^{nR_{i}} \}$, $i=1,2,\ldots,K$. The decoding function of receiver $i$ is
$\hat{W}_{i}(\y_{i}^{n})$ . An error occurs when $\hat{W}_{i} \ne W_{i}$. A rate vector $\R =
[R_{1}, \ldots, R_{K}]^{T}$ is said to be achievable if there exists a sequence of $(n, 2^{nR_{1}},
\ldots, 2^{nR_{K}})_{\mathrm{BC}}$ codebooks for which the average probability of error $P_{e}
\rightarrow 0$ as the code length $n \rightarrow \infty$. The capacity region of a broadcast
channel is defined as the union of all achievable rate vectors \cite{Cover91:Info_Thry}. Gaussian
vector broadcast channel can be used to model many different types of systems
\cite{Cover91:Info_Thry}. Due to the close relationship between Gaussian vector broadcast channel
and MIMO, we will call the Gaussian vector broadcast channel in the MIMO case as MIMO-BC throughout
the rest of this paper.

For clarity, we use $\GG_{i}$ to specifically denote the input covariance matrix of link $i$ in a
MIMO-BC, and $\Q_{j}$ to denote an input covariance matrix in other types of MIMO channels. From
the encoding process of DPC, the achievable rate in DPC scheme can be computed as follows:
\begin{equation} \label{eqn_dpc_link_rate}
R_{\pi(i)} = \log \frac{\ddet {\I + \Hc_{\pi(i)} \left( \sum_{j \geq i} \GG_{\pi(j)} \right)
\Hc_{\pi(i)}^{\dag} } }{ \ddet { \I + \Hc_{\pi(i)} \left( \sum_{j > i} \GG_{\pi(j)} \right)
\Hc_{\pi(i)}^{\dag} } },
\end{equation}
where $\pi$ denotes a permutation of the set $\{1,\ldots,K\}$, $\pi(i)$ represents the $i^{th}$
position in permutation $\pi$. One important observation of the dirty paper rate equation in
(\ref{eqn_dpc_link_rate}) is that the rate equation is neither a concave nor a convex function of
the input covariance matrices $\GG_{i}$, $i=1,2,\ldots,K$.

Let $\Hc = [\Hc_{1}, \ldots, \Hc_{K}]^{T}$ be the collection of $K$ channel gain matrices in the
MIMO-BC, and $\GG = [\GG_{1}\,\ldots,\GG_{K}]$ be the collection of $K$ input covariance matrices.
We define the dirty paper region $\mathcal{C}_{DPC}(P, \Hc)$ as the convex hull of the union of all
such rates vectors over all positive semidefinite covariance matrices $\GG_{1}, \ldots, \GG_{K}$
satisfying $\tr\{ \sum_{i=1}^{K} \GG_{i} \} \leq P$ (the maximum transmit power constraint at the
transmitter) and over all $K!$ permutations:
\begin{equation*}
\mathcal{C}_{\mathrm{DPC}}(P,\Hc) \triangleq Cov\left( \cup_{\pi,\GG} \R^{\mathrm{BC}}(\pi, \GG)
\right),
\end{equation*}
where $Cov(\cdot)$ represents the convex hull operation.

\subsection{Problem Formulation}
In this paper, we aim to solve the problem of jointly optimizing DPC per-antenna power allocation
at each node in the link layer and multihop/multipath routing in a MIMO-based WMN. Suppose that
each node in the network has been assigned a certain (possibly reused) frequency band that will not
cause interference to any other node in the network. Also, the incoming and outgoing bands of each
node are non-overlapping such that each node can transmit and receive simultaneously. How to
perform channel assignments is a huge research topic on its own merits, and there are a vast amount
of literature that discuss channel assignment problems. Thus, in this paper, we focus on how to
jointly optimize routing in the network layer and the DPC power allocation in the link layer for
each node when a channel assignment is given. We adopt the well-known proportional fairness utility
function, i.e., $\ln(s_{f})$ for flow $f$. In CRPA, we wish to maximize the sum of all utility
functions. In the link layer, since the total transmit power of each node is subject to a maximum
power constraint, we have $\sum_{l \in \Out{n}} \tr\{ \GG_{l} \} \leq P_{\max}^{(n)}$, $1 \leq n
\leq N$, where $P_{\max}^{(n)}$ represents the maximum transmit power of node $n$. Since the total
amount of flow in each link $l$ cannot exceed its capacity limit, we must have $\sum_{f=1}^{F}
t_{l}^{(f)} \leq R_{l}(\GG)$, $1 \leq l \leq L$. This can be further compactly written using
matrix-vector notations as $\langle \1, \T^{T}\e_{l} \rangle \leq R_{l}(\GG)$, $1 \leq l \leq L$.
Coupling the network layer model in Section~\ref{sec:network_flow} and MIMO-BC link layer model in
Section~\ref{sec:mimo_linkcap}, we have the problem formulation for CRPA as in (\ref{eqn_crpa}).
\begin{equation} \label{eqn_crpa}
\begin{array}{rll}
\mbox{\underline{\textbf{CRPA:}}}& & \\
\mbox{Maximize} & \sum_{f=1}^{F} \ln(s_{f}) & \\
\mbox{subject to} & \A \T = \S & \\
& \T \geq \0 & \\
& \S \e_{f} =_{\src{f},\dst{f}} \0 & \!\!\! \forall \, f \\
& \langle \1, \S\e_{f} \rangle = 0 & \!\!\! \forall \, f \\
& (\S \e_{f})_{\src{f}} = s_{f} & \!\!\! \forall \, f \\
& \langle \1 , \T^{T}\e_{l} \rangle \leq R_{l}(\GG) & \!\!\! \forall \, l \\
& R_{l}(\GG) \in \mathcal{C}_{\mathrm{DPC}}^{(n)}(P_{\max}^{(n)}, \Hc^{(n)}) & \!\!\! \forall l \in \Out{n} \\
& \sum_{l \in \Out{n}} \tr\{ \GG_{l} \} \leq P_{\max}^{(n)} & \!\!\! \forall \, n \\
& \GG_{l} \succeq 0 & \!\!\! \forall \, l \\
& \mbox{Variables: } \S, \, \T, \, \GG & \\
\end{array}
\end{equation}

\section{Reformulation of CRPA}\label{sec:reformulation}
As we pointed out earlier, the DPC rate equation in (\ref{eqn_dpc_link_rate}) is neither a concave
nor a convex function of the input covariance matrices. As a result, the cross-layer optimization
problem in (\ref{eqn_crpa}) is a non-convex optimization problem, which is very hard to solve
numerically, let alone analytically. However, in the following, we will show that (\ref{eqn_crpa})
can be reformulated as an equivalent convex optimization problem by projecting all the MIMO-BC
channels onto their dual MIMO multiple-access channels (MIMO-MAC). We first provide some background
of Gaussian vector multiple access channels and the channel duality between MIMO-BC and MIMO-MAC.

\subsection{MIMO-MAC Channel Model}
\begin{figure}[ht!]
\centering
\includegraphics[width=3.0in]{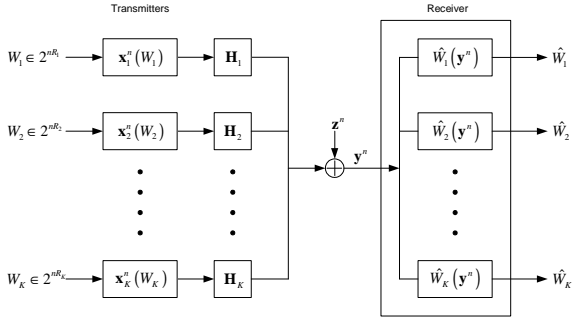}
\caption{A Gaussian vector multiple access channel.} \label{fig_mac}
\end{figure}

A communication system where multiple {\em uncoordinated} transmitters send independent information
to a single receiver is referred to as a multiple access channel. If the channel gain of each link
in the multiple access channel is a matrix and the noise is a Gaussian random vector, the channel
is termed ``Gaussian vector multiple access channel''. Fig.~\ref{fig_mac} illustrates a $K$-user
Gaussian vector multiple access channel, where independent messages $W_{1},\ldots, W_{K}$, are
encoded by transmitters 1 to $K$, respectively, and the receiver is trying to decode $W_{1},\ldots,
W_{K}$. A $(n, 2^{nR_{1}}, \ldots, 2^{nR_{K}})_{\mathrm{MAC}}$ codebook for a multiple access
channel consists of encoding functions $\x_{1}^{n}(W_{1}), \ldots, \x_{K}^{n}(W_{K})$ where $W_{i}
\in \{ 1,\ldots, 2^{nR_{i}} \}$, $i=1,2,\ldots,K$. The decoding functions at the receiver are
$\hat{W}_{i}(\y_{i}^{n})$, $i=1,2,\ldots,K$. An error occurs when $\hat{W}_{i} \ne W_{i}$. A rate
vector $\R = [R_{1}, \ldots, R_{K}]^{T}$ is said to be achievable if there exists a sequence of
$(n, 2^{nR_{1}}, \ldots, 2^{nR_{K}})_{\mathrm{MAC}}$ codebooks for which the average probability of
error $P_{e} \rightarrow 0$ as the code length $n \rightarrow \infty$. The capacity region of a
multiple access channel, denoted by $\Cmac$ is defined as the union of all achievable rate vectors
\cite{Cover91:Info_Thry}. We call the Gaussian vector multiple access channel in the MIMO case as
MIMO-MAC throughout the rest of this paper.

\subsection{Duality between MIMO-BC and MIMO-MAC}
\begin{figure}[ht!]
\centering
\includegraphics[width=3.5in]{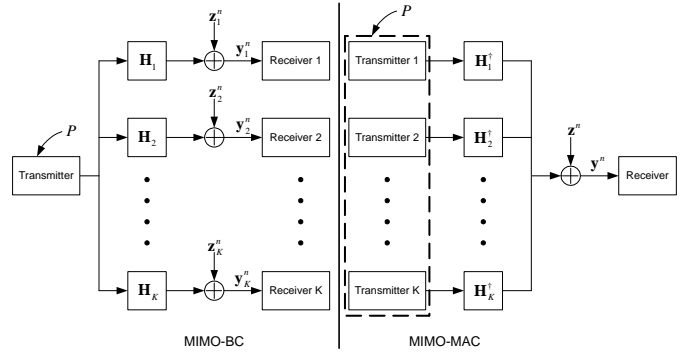}
\caption{The relationship between MIMO-BC and its dual MIMO-MAC} \label{fig_bc_mac}
\end{figure}

The dual MIMO-MAC of a MIMO-BC can be constructed by changing the receivers in the MIMO-BC into
transmitters and changing the transmitter in the MIMO-BC into the receiver. The channel gain
matrices in dual MIMO-MAC are the conjugate transpose of the channel gain matrices in MIMO-BC. The
maximum sum power in the dual MIMO-MAC is the same maximum power level as in MIMO-BC. The
relationship between a MIMO-BC and its dual MIMO-MAC is illustrated in Fig.~\ref{fig_bc_mac}.
Similar to MIMO-BC, We denote the capacity region of the dual MIMO-MAC as $\Cmac(P, \Hc^{\dag})$.
The following Lemma states the relationship between the capacity regions of a MIMO-BC and its dual
MIMO-MAC.
\begin{lem}\label{thm_mac_bc_dual} The DPC region of a MIMO-BC channel with maximum power constraint $P$ is
equal to the capacity region of the dual MIMO MAC with sum power constraint $P$
\begin{equation*}
\Cbc(P, \Hc) = \Cmac (P, \Hc^{\dag}).
\end{equation*}
\end{lem}
\begin{proof}
The proof of this theorem can be arrived in various ways \cite{Vishwanath03:duality,
Viswanath03:duality, Yu06:duality}. The most straightforward approach is to show that any MIMO-BC
achievable rate vector is also achievable in its dual MIMO-MAC and vice versa. The MAC-to-BC and
BC-to-MAC mappings can be found in \cite{Vishwanath03:duality}. It is also shown in
\cite{Vishwanath03:duality} that any rate vector in a MIMO-BC with a particular encoding order can
be achieved in its dual MIMO-MAC with the {\em reversed} successive decoding order.
\end{proof}

\subsection{Convexity of MIMO-MAC Capacity Region}
From Lemma~\ref{thm_mac_bc_dual}, we know that the capacity region of a MIMO-BC and its dual
MIMO-MAC is exactly the same. Therefore, we can replace $\Cbc(\cdot)$ in (\ref{eqn_crpa}) by the
capacity regions of the dual MIMO-MAC channels $\Cmac(\cdot)$. The benefits of such replacements is
due to the following theorem.
\begin{thm} \label{thm_mac_convex}
The capacity region of a $K$-user MIMO-MAC channel with a sum power constraint $\sum_{i=1}^{K}
\tr(\Q_{i}) \leq P_{\max}$ is convex with respect to the input covariance matrices
$\Q_{1},\ldots,\Q_{K}$.
\end{thm}
\begin{proof}
Denote the input signals of the $K$ users by $\x_{1},\ldots,\x_{K}$, respectively, and denote the
output of the MIMO-MAC channel by $\y$. Since $\rho_{i}$ is a scalar, we absorb $\rho_{i}$ into
$\Hc_{i}$ in this proof for notation convenience. Theorem 14.3.5 in \cite{Cover91:Info_Thry} states
that the capacity region of a MIMO-MAC is determined by
\begin{eqnarray} \label{eqn_thm2_1}
&& \hspace{-.5in} \Cmac(\Q_{1},\ldots,\Q_{K})= \nonumber\\
&& \hspace{-.5in} Cov \left\{ \hspace{-.0in} (R_{1},\ldots,R_{K}) \left|
\begin{array}{l}
\hspace{-.05in} \sum_{i \in \mathcal{S}} R_{i}(\Q) \leq \\
\hspace{-.05in} I(\x_{i}, i \in \mathcal{S}; \y| \x_{i}, {i \in \mathcal{S}^{c}} ), \\
\hspace{-.05in} \forall \mathcal{S} \subseteq \{1,\ldots,K\} \\
\hspace{-.05in} \sum_{i=1}^{K} \tr(\Q_{i}) \leq P_{\max}
\end{array}
\right. \hspace{-.1in}\right\},
\end{eqnarray}
where the mutual information expression $I(;)$ can be bounded as follows:
\begin{equation} \label{eqn_mutual_info_bnd}
\hspace{0in} I(\x_{i}, {i \in \mathcal{S}}; \y | \x_{i}, {i \in \mathcal{S}^{c}}) \leq \log
\ddet{\I + \sum_{i\in \mathcal{S}} \Hc_{i}^{\dag} \Q_{i} \Hc_{i}}.
\end{equation}
To show that the capacity region of the MIMO-MAC with a sum power constraint is convex, it is
equivalent to show that the convex hull operation in (\ref{eqn_thm2_1}) is unnecessary. To show
this, consider the convex combination of two arbitrarily chosen achievable rate vectors
$[R_{1},\ldots, R_{K}]$ and $[\hat{R}_{1}, \ldots, \hat{R}_{K}]$ determined by two feasible power
vectors $[\Q_{1}, \ldots, \Q_{K}]$ and $[\hat{\Q}_{1},\ldots,\hat{\Q}_{K}]$, respectively, i.e., we
have $\sum_{i=1}^{K} \tr(\Q_{i}) \leq P_{\max}$ and $\sum_{i=1}^{K} \tr(\hat{\Q}_{i}) \leq
P_{\max}$. Let $0 \leq \alpha \leq 1$ and consider the convex combination
\begin{equation*}
[\bar{R}_{1},\ldots,\bar{R}_{K}] = \alpha [R_{1},\ldots, R_{K}] + (1-\alpha) [\hat{R}_{1}, \ldots,
\hat{R}_{K}].
\end{equation*}
Also, let $\bar{\Q}_{i} = \alpha \Q_{i} + (1-\alpha) \hat{\Q}_{i}$, $i=1,\ldots,K$. It is easy to
verify that $\sum_{i=1}^{K} \tr(\bar{\Q}_{i}) \leq P_{\max}$, i.e., the convex combination of two
feasible power vectors  is also feasible. Now, consider
\begin{eqnarray*}
&& \hspace{-.3in}\alpha \sum_{i\in \mathcal{S}} R_{i} + (1-\alpha) \sum_{i \in \mathcal{S}}
\hat{R}_{i} \leq \alpha \log \ddet{\I + \sum_{i\in \mathcal{S}} \Hc_{i}^{\dag} \Q_{i}
\Hc_{i} } \\
&& + (1-\alpha) \log \ddet{\I + \sum_{i\in \mathcal{S}} \Hc_{i}^{\dag} \hat{\Q}_{i} \Hc_{i} }.
\end{eqnarray*}
Since the function $\log |\A|$ is a concave function for any positive semidefinite matrix variable
$\A$ \cite{Cover91:Info_Thry}, it follows from Jensen's inequality that
\begin{equation*}
\alpha \sum_{i\in \mathcal{S}} R_{i} + (1-\alpha) \sum_{i \in \mathcal{S}} \hat{R}_{i} \leq
\frac{1}{2} \log \ddet{\I + \sum_{i\in \mathcal{S}} \Hc_{i}^{\dag} \bar{\Q}_{i} \Hc_{i} },
\end{equation*}
which means that the convex combination of rate vectors $[R_{1},\ldots, R_{K}]$ and $[\hat{R}_{1},
\ldots, \hat{R}_{K}]$ can also be achieved by using the feasible power vector
$[\bar{\Q}_{1},\ldots,\bar{\Q}_{K}]$ directly. As a result, the convex hull operation is
unnecessary.
\end{proof}

\subsection{Maximum Weighted Sum Rate Problem of the Dual MIMO-MAC}
Now, we consider the maximum weighted sum rate problem of the dual MIMO-MAC. We simplify this
problem such that we do not have to enumerate all possible successive decoding order in the dual
MIMO-MAC, thus paving the way to efficiently solve the link layer subproblem we discuss in
Section~\ref{sec:soln_proc}.
\begin{thm} \label{thm_weighted_sum}
Associate each rate $R_{i}$ in MIMO-MAC a non-negative weight $u_{i}$, $i=1,\ldots,K$, the maximum
weighted sum $\max \sum_{i=1}^{K} R_{i}(\Q)$ can be solved by the following convex optimization
problem:
\begin{equation} \label{eqn_thm3_1}
\begin{array}{rl}
\!\!\!\!\!\! \mbox{Maximize} & \!\!\! \sum_{i=1}^{K} (u_{\pi(i)} - u_{\pi(i-1)}) \times \\
\!\!\!\!\!\! & \!\!\! \log \ddet{\I + \sum_{j=i}^{K} \rho_{\pi(j)} \Hc_{\pi(j)}^{\dag} \Q_{\pi(j)} \Hc_{\pi(j)} } \\
\!\!\!\!\!\!  \mbox{subject to} & \!\!\! \sum_{i=1}^{K} \tr(\Q_{i}) \leq P_{\max} \\
\!\!\!\!\!\! & \!\!\! \Q_{i} \succeq 0, \,\, i=1,\ldots,K,
\end{array}
\end{equation}
where $u_{\pi(0)} \triangleq 0$, $\pi(i), i=1,\ldots,K$ is a permutation on $\{1,\ldots,K\}$ such
that $u_{\pi(1)} \leq \ldots \leq u_{\pi(K)}$. In particular, suppose that
$(\Q_{\pi(1)}^{*},\ldots,\Q_{\pi(K)}^{*})$ solves (\ref{eqn_thm3_1}), then the optimal rates of
(\ref{eqn_thm3_1}) are given by
\begin{equation}
R_{\pi(K)}^{*} = \log \ddet {\I + \rho_{\pi(K)} \Hc_{\pi(K)}^{\dag} \Q_{\pi(K)}^{*} \Hc_{K} }
\end{equation}
and
\begin{eqnarray}
\hspace{-.2in} & R_{\pi(i)}^{*} = \log \ddet {\I + \sum_{j=i}^{K} \rho_{\pi(j)} \Hc_{\pi(j)}^{\dag}
\Q_{\pi(j)}^{*} \Hc_{j} }& \nonumber\\
\hspace{-.2in} & \quad \quad - \log \ddet {\I + \sum_{j=i+1}^{K} \rho_{\pi(j)} \Hc_{\pi(j)}^{\dag}
\Q_{\pi(j)}^{*} \Hc_{j} }, &
\end{eqnarray}
for $i=1,2,\ldots,K-1$.
\end{thm}

\begin{proof}
For convenience, we let
\begin{equation*}
\Phi(\mathcal{S}) = \log \ddet{\I + \sum_{i\in \mathcal{S}} \Hc_{\pi(i)}^{\dag} \Q_{\pi(i)}
\Hc_{\pi(i)}}.
\end{equation*}
Since $\pi(i)$ is simply a permutation on $\{1,\ldots,K\}$, from (\ref{eqn_thm2_1}) and
(\ref{eqn_mutual_info_bnd}) we have the maximum weighted sum rate problem can be written as
\begin{equation*}
\begin{array}{rl}
\mbox{Maximize} & \sum_{i=1}^{K} u_{\pi(i)} R_{\pi(i)} \\
\mbox{subject to} & \sum_{i \in \mathcal{S}} R_{\pi(i)} \leq \Phi(\mathcal{S}), \,\, \forall
\mathcal{S} \subseteq \{1,\ldots,K\}.
\end{array}
\end{equation*}
Also from (\ref{eqn_thm2_1}) and (\ref{eqn_mutual_info_bnd}), it is easy to derive that $R_{\pi(i)}
\leq \Phi(\{\pi(i)\}) = \log\ddet{\I + \rho_{\pi(i)} \Hc_{\pi(i)}^{\dag} \Q_{\pi(i)} \Hc_{\pi(i)}
}$. Since $u_{\pi(1)} \leq \ldots \leq u_{\pi(K)}$, from KKT condition, we must have that the
constraint $R_{\pi(K)} = \Phi(\{ \pi(K) \})$ must be tight at optimality. That is,
\begin{equation} \label{eqn_R_K}
R_{\pi(K)} = \log \ddet {\I + \rho_{\pi(K)} \Hc_{\pi(K)}^{\dag} \Q_{\pi(K)} \Hc_{\pi(K)} }.
\end{equation}
Again, from (\ref{eqn_thm2_1}) and (\ref{eqn_mutual_info_bnd}), we have
\begin{eqnarray}
\hspace{-.2in} & R_{\pi(K-1)} + R_{\pi(K)} \leq \log \left| \I + \rho_{\pi(K)} \Hc_{\pi(K)}^{\dag}
\Q_{\pi(K)} \Hc_{\pi(K)} \right. \nonumber\\
\hspace{-.2in} & \left.+ \rho_{\pi(K-1)} \Hc_{\pi(K-1)}^{\dag} \Q_{\pi(K-1)} \Hc_{\pi(K-1)}
\right|. & \nonumber
\end{eqnarray}
So,
\begin{eqnarray} \label{eqn_thm3_2}
& R_{\pi(K-1)} \leq \log \left| \I + \rho_{\pi(K)} \Hc_{\pi(K)} \Q_{\pi(K)}
\Hc_{\pi(K)}^{\dag} \right. & \nonumber\\
& \left. + \rho_{\pi(K-1)} \Hc_{\pi(K-1)} \Q_{\pi(K-1)} \Hc_{\pi(K-1)}^{\dag} \right| - \nonumber\\
& \log \ddet{\I + \rho_{\pi(K)} \Hc_{\pi(K)} \Q_{\pi(K)} \Hc_{\pi(K)}^{\dag} } &
\end{eqnarray}
Since $u_{\pi(K-1)}$ is the second largest weight, again from KKT condition, we must have that
(\ref{eqn_thm3_2}) must be tight at optimality. This process continues for all $K$ users.
Subsequently, we have that
\begin{eqnarray} \label{eqn_thm3_3}
&R_{\pi(i)} = \log \ddet{\I + \sum_{j=i}^{K} \rho_{\pi(j)} \Hc_{\pi(j)}^{\dag}
\Q_{\pi(j)} \Hc_{j} }& \nonumber\\
& \quad \quad - \log \ddet{\I + \sum_{j=i+1}^{K} \rho_{\pi(j)} \Hc_{\pi(j)}^{\dag} \Q_{\pi(j)}
\Hc_{j} }, &
\end{eqnarray}
for $i=1,\ldots,K-1$. Summing up all $u_{\pi(i)} R_{\pi(i)}$ and after rearranging the terms, it is
readily verifiable that
\begin{eqnarray} \label{eqn_thm3_4}
&& \hspace{-.2in} \sum_{i=1}^{K} u_{\pi(i)} R_{\pi(i)} = \sum_{i=1}^{K} (u_{\pi(i)} - u_{\pi(i-1)}) \times \nonumber \\
&& \hspace{-.1in} \log \ddet{\I + \sum_{j=i}^{K} \rho_{\pi(j)} \Hc_{\pi(j)}^{\dag} \Q_{\pi(j)}
\Hc_{\pi(j)} }.
\end{eqnarray}
It then follows that the maximum weighted sum rate problem of MIMO-MAC is equivalent to maximizing
(\ref{eqn_thm3_4}) subject to the sum power constraint, i.e., the optimization problem in
(\ref{eqn_thm3_1}). Since $\log\ddet{\cdot}$ is a concave function of positive semidefinite
matrices, (\ref{eqn_thm3_4}) is a convex optimization problem with respect to
$\Q_{\pi(1)},\ldots,\Q_{\pi(K)}$. After we obtain the optimal solution power solution
$(\Q_{\pi(1)}^{*},\ldots,\Q_{\pi(K)}^{*})$, the corresponding link rates can be computed by simply
following (\ref{eqn_R_K}) and (\ref{eqn_thm3_3}).
\end{proof}

\subsection{Problem Reformulation}
We now reformulate CRPA by replacing $\Cbc$ in (\ref{eqn_crpa}) with $\Cmac$, and we denote the
equivalent problem by $\mbox{CRPA-E}$. After solving CRPA-E, we can recover the corresponding
MIMO-BC covariance matrices $\GG^{*}$ from the optimal solution $\Q^{*}$ of CRPA-E by the MAC-to-BC
mapping provided in \cite{Vishwanath03:duality}.
\begin{equation} \label{eqn_crpa_equiv}
\begin{array}{rll}
\hspace{-.1in} \mbox{\underline{\textbf{CRPA-E:}}} & & \\
\hspace{-.1in} \mbox{Maximize} & \sum_{f=1}^{F} \ln(s_{f}) & \\
\hspace{-.1in} \mbox{subject to} & \A \T = \S & \\
\hspace{-.1in} & \T \geq \0 & \\
\hspace{-.1in} & \S \e_{f} =_{\src{f},\dst{f}} \0 &  \!\!\! \forall \, f \\
\hspace{-.1in} & \langle \1, \S\e_{f} \rangle = 0 &  \!\!\! \forall \, f \\
\hspace{-.1in} & (\S \e_{f})_{\src{f}} = s_{f} &  \!\!\! \forall \, f \\
\hspace{-.1in} & \langle \1 , \T^{T}\e_{l} \rangle \leq R_{l}(\Q) & \!\!\! \forall \, l \\
\hspace{-.1in} & R_{l}(\Q) \in \mathcal{C}_{\mathrm{MAC}}^{(n)}(P_{\max}^{(n)}, \Hc^{\dag(n)}) &  \!\!\! \forall l \in \Out{n} \\
\hspace{-.1in} & \sum_{l \in \Out{n}} \tr\{ \Q_{l} \} \leq P_{\max}^{(n)} &  \!\!\! \forall \, n \\
\hspace{-.1in} & \Q_{l} \succeq 0 &  \!\!\! \forall \, l \\
\hspace{-.1in} & \mbox{Variables: } \S, \, \T, \, \Q& \\
\end{array}
\end{equation}

\section{Solution Procedure} \label{sec:soln_proc}
Since CRPA-E is a convex programming problem, we can solve CRPA-E exactly by solving its Lagrangian
dual problem. Introducing Lagrangian multipliers $u_{i}$ to the link capacity coupling constraints
$\langle \1 , \T^{T}\e_{l} \rangle \leq R_{l}(\Q)$, Hence, we can write the Lagrangian as
\begin{equation} \label{eqn_lagrangian}
\Theta(\u) = \sup_{\S,\T,\Q} \left\{ L(\S,\T,\Q,\u) | (\S, \T, \Q) \in \Psi \right\},
\end{equation}
where
\begin{equation*}
L(\S,\T,\Q,\u) = \sum_{f} \ln \left( s_{f} \right) + \sum_{l} u_{l} \left(R_{l}(\Q) - \langle \1,
\T^{T}\e_{l} \rangle \right)
\end{equation*}
and $\Psi$ is defined as
\begin{equation*}{\small
\Psi \triangleq \left\{ (\S, \T, \Q) \left|
\begin{array}{ll}
\A \T = \S \\
\T \geq \0 \\
\S \e_{f} =_{\src{f},\dst{f}} \0 & \forall \, f \\
\langle \1, \S\e_{f} \rangle = 0 & \forall \, f \\
(\S \e_{f})_{\src{f}} = s_{f} & \forall \, f \\
\sum_{l \in \Out{n}} \tr\{ \Q_{l} \} \leq P_{\max}^{(n)} & \forall \, n \\
\Q_{l} \succeq 0 & \forall \, l \\
R_{l}(\Q) \in \mathcal{C}_{\mathrm{MAC}}(P_{\max}^{(n)}, \Hc^{\dag(n)}) & \forall \, n \\
\end{array} \right.
\right\} }.
\end{equation*}
The Lagrangian dual problem of CRPA can thus be written as:
\begin{equation*}
\begin{array}{rll}
\mathbf{D}^{\mathrm{CRPA-E}}: \quad \mbox{Minimize} & \Theta(\u) & \\
\mbox{subject to} & \u \geq \0. \\
\end{array}
\end{equation*}
It is easy to recognize that, for a given $\u$, the Lagrangian in (\ref{eqn_lagrangian}) can be
rearranged and separated into two terms:
\begin{equation*}
\Theta(\u) = \Theta_{\net}(\u) + \Theta_{\link}(\u),
\end{equation*}
where, for a given Lagrangian multiplier $\u$, $\Theta_{\net}$ and $\Theta_{\link}$ are
corresponding to network layer and link layer variables, respectively:
\begin{eqnarray*}
&& \begin{array}{rll}
\hspace{-.2in}\mathbf{D}^{\mathrm{CRPA-E}}_{\net}:
\Theta_{\net}(\u)\triangleq &
\mbox{Maximize} \sum_{f} \ln \left(s_{f} \right) \\
\hspace{-.2in}& \quad - \sum_{l} u_{l} \langle \1, \T^{T}\e_{l} \rangle & \\
\hspace{-.2in}\mbox{subject to} & \A \T = \S & \\
\hspace{-.2in}& \T \geq \0 & \\
\hspace{-.2in}& \S \e_{f} =_{\src{f},\dst{f}} \0 & \forall \, f \\
\hspace{-.2in}& \langle \1, \S\e_{f} \rangle = 0 & \forall \, f \\
\hspace{-.2in}& (\S \e_{f})_{\src{f}} = s_{f} & \forall \, f \\
\hspace{-.2in}\mbox{Variables: } & \S, \, \T &
\end{array}
\end{eqnarray*}
\begin{eqnarray*}
&& \begin{array}{rll}
\hspace{-.35in} \mathbf{D}^{\mathrm{CRPA-E}}_{\link}: \Theta_{\link}(\u) \triangleq & \mbox{Maximize} \sum_{l} u_{l} R_{l}(\Q) & \\
\hspace{-.4in} \mbox{subject to} & \sum_{l \in \Out{n}} \tr\{ \Q_{l} \} \leq P_{\max}^{(n)} & \hspace{-.2in} \forall \, n \\
\hspace{-.4in} & \Q_{l} \succeq 0 & \hspace{-.2in} \forall \, l \\
\hspace{-.4in} & R_{l}(\Q) \in \mathcal{C}_{\mathrm{MAC}}(P_{\max}^{(n)}, \Hc^{\dag(n)}), & \\
\hspace{-.3in} & \quad \quad \forall \, l \in \Out{n}, n \in N & \\
\hspace{-.4in} \mbox{Variables: } & \Q&
\end{array}
\end{eqnarray*}

The CRPA-E Lagrangian dual problem can thus be written as the following master dual problem:
\begin{equation*}
\begin{array}{rll}
\mathbf{MD}^{\mathrm{CRPA-E}}: \quad \mbox{Minimize} & \Theta_{\net}(\u) + \Theta_{\link}(\u) & \\
\mbox{subject to} & \u \geq \0 \\
\end{array}
\end{equation*}
Notice that $\Theta_{\link}(\u)$ can be further decomposed on a node-by-node basis as follows:
\begin{eqnarray}
&& \hspace{-.4in} \Theta_{\link}(\u) = \max \sum_{l} u_{l} R_{l}(\Q) \nonumber\\
&& \hspace{-.4in} = \sum_{n=1}^{N} \left( \max \sum_{l \in \Out{n}} u_{l} R_{l}(\Q) \right)  =
\sum_{n=1}^{N} \Theta_{\link}^{(n)}(\u^{(n)}).
\end{eqnarray}
It is seen that $\Theta_{\link}^{(n)} (\u^{(n)}) \triangleq \max \sum_{l \in \Out{n}} u_{l}
R_{l}(\Q)$ is a maximum weighted sum rate problem of the dual MIMO-MAC for some given dual
variables $\u^{(n)}$ as weights. Without loss of generality, suppose that node $n$ has $K$ outgoing
links, which are indexed as $1,\ldots,K$ and are associated with dual variables $u_{1}, \ldots,
u_{K}$, respectively. Let $\pi(i) \in \{1,\ldots,K\}$ be the permutation such that $0 \leq
u_{\pi(1)} \leq \ldots \leq u_{\pi(K)}$ and define $u_{\pi(0)}=0$. $\Theta_{\link}^{(n)}
(\u^{(n)})$ can be written as follows:
\begin{equation} \label{eqn_bc_node}
\begin{array}{rl}
\hspace{-.1in} \mbox{Maximize} & \!\!\! \sum_{i=1}^{K} (u_{\pi(i)} - u_{\pi(i-1)}) \times \\
\hspace{-.1in} & \!\!\! \log \ddet{\I + \sum_{j=i}^{K} \rho_{\pi(j)} \Hc_{\pi(j)}^{\dag} \Q_{\pi(j)} \Hc_{\pi(j)} } \\
\hspace{-.1in}  \mbox{subject to} & \!\!\! \sum_{i=1}^{K} \tr(\Q_{i}) \leq P_{\max}^{(n)} \\
\hspace{-.1in} & \!\!\! \Q_{i} \succeq 0, \,\, i=1,\ldots,K.
\end{array}
\end{equation}
Note that in the network layer subproblem $\Theta_{\net}(\u)$, the objective function is concave
and all constraints are affine. Therefore, $\Theta_{\net}(\u)$ is readily solvable by many
polynomial time convex programming methods. However, even though $\Theta_{\link}^{(n)}(\u^{(n)})$
is also a convex problem, generic convex programming methods are not efficient because the
structures of its objective function and constraints are very complex. In the following
subsections, we will discuss in detail how to solve $\Theta_{\link}^{(n)}(\u^{(n)})$.

\subsection{Conjugate Gradient Projection for Solving $\Theta_{\link}^{(n)}(\u^{(n)})$} \label{sec:framework}
We propose an efficient algorithm based on conjugate gradient projection (CGP) to solve
(\ref{eqn_bc_node}). CGP utilizes the important and powerful concept of Hessian conjugacy to
deflect the gradient direction appropriately so as to achieve the superlinear convergence rate
\cite{Bazaraa_Sherali_Shetty_93:NLP}, which is similar to that of the well-known quasi-Newton
methods (e.g., BFGS method). In each iteration, CGP projects the conjugate gradient direction to
find an improving feasible direction. The framework of CGP for solving (\ref{eqn_bc_node}) is shown
in Algorithm~\ref{alg_mgp}.
\begin{algorithm}
{\footnotesize \caption{Gradient Projection Method} \label{alg_mgp}
\begin{algorithmic}
\STATE {\bf Initialization:} \\
\STATE \quad Choose the initial conditions $\Q^{(0)} = [ \Q_{1}^{(0)}, \Q_{2}^{(0)}, \ldots, \Q_{K}^{(0)}]^{T}$. Let \\
\STATE \quad $k=0$. \\
\STATE {\bf Main Loop:} \\
\STATE \quad 1. Calculate the conjugate gradients $\G_{i}^{(k)}$, $i= 1,2, \ldots, K$. \\
\STATE \quad 2. Choose an appropriate step size $s_{k}$. Let $\Q_{i}^{'(k)} = \Q_{i}^{(k)} + s_{k}
\G_{i}^{(k)}$, \\
\STATE \quad \quad for $i=1,2,\ldots,K$. \\
\STATE \quad 3. Let $\bar{\Q}^{(k)}$ be the projection of $\Q^{'(k)}$ onto $\Pwrcon$. \\
\STATE \quad 4. Choose an appropriate step size $\alpha_{k}$. Let $\Q_{l}^{(k+1)} = \Q_{l}^{(k)} +$ \\
\STATE \quad \quad $\alpha_{k}(\bar{\Q}_{i}^{(k)} - \Q_{i}^{(k)})$, $i=1,2,\ldots,K$. \\
\STATE \quad 5. $k=k+1$. If the maximum absolute value of the elements in $\Q_{i}^{(k)} - $\\
\STATE \quad \quad $\Q_{i}^{(k-1)}< \epsilon$, for $i=1,2,\ldots,L$, then stop; else go to step 1.
\end{algorithmic}}
\end{algorithm}

We adopt the ``Armijo's Rule'' inexact line search method to avoid excessive objective function
evaluations, while still enjoying provable convergence \cite{Bazaraa_Sherali_Shetty_93:NLP}. For
convenience, we use $F(\Q)$ to represent the objective function in (\ref{eqn_bc_node}), where $\Q =
(\Q_{1},\ldots,\Q_{K})$ denotes the set of covariance matrices at a node. According to Armijo's
Rule, in the $k^{th}$ iteration, we choose $\sigma_{k}=1$ and $\alpha_{k} = \beta^{m_{k}}$ (the
same as in \cite{Ye03:MIMO_SH_AdHoc}), where $m_{k}$ is the first non-negative integer $m$ that
satisfies
\begin{eqnarray} \label{eqn_Armijo}
\!\!\!\!\!\! && F(\Q^{(k+1)}) - F(\Q^{(k)})\geq \sigma \beta^{m} \langle
\G^{(k)}, \bar{\Q}^{(k)} - \Q^{(k)} \rangle \nonumber\\
\!\!\!\!\!\! && = \sigma \beta^{m} \sum_{i=1}^{K} \tr \left[ \G_{i}^{\dag(k)}
\left(\bar{\Q}_{i}^{(k)} - \Q_{i}^{(k)}\right) \right],
\end{eqnarray}
where $0 < \beta < 1$ and $0 < \sigma < 1$ are fixed scalars.

\subsection{Computing the Conjugate Gradients} The gradient $\bar{\G}_{\pi(j)} \triangleq \nabla_{\Q_{\pi(j)}}
F(\Q)$ depends on the partial derivatives of $F(\Q)$ with respect to $\Q_{\pi(j)}$. By using the
formula $\frac{\partial \ln \ddet{\A+\B\X\C}}{\partial \X} = \left[ \C(\A+\B\X\C)^{-1}\B
\right]^{T}$ \cite{Ye03:MIMO_SH_AdHoc, Magnus_Neudecker99:Mtrx_Diff_Calc}, we can compute the
partial derivative of the $i^{th}$ term in the summation of $F(\Q)$ with respect to $\Q_{\pi(j)}$,
$j \geq i$, as follows:
\begin{eqnarray*} \label{eqn_partial}
&& \hspace{-.25in}\frac{\partial} {\partial \Q_{\pi(j)}} \Bigg( (u_{\pi(i)} - u_{\pi(i-1)}) \times
\nonumber \\
&& \hspace{.3in} \left. \log \ddet { \I + \sum_{k=i}^{K} \rho_{\pi(k)} \Hc_{\pi(k)}^{\dag} \Q_{\pi(k)} \Hc_{\pi(k)} } \right) \nonumber \\
&& \hspace{-.25in} = \rho_{\pi(j)} \left(u_{\pi(i)} - u_{\pi(i-1)}\right) \times \nonumber\\ &&
\hspace{-.3in} \left[ \Hc_{\pi(j)} \left( \I + \sum_{k=i}^{K} \rho_{\pi(k)} \Hc_{\pi(k)}^{\dag}
\Q_{\pi(k)} \Hc_{\pi(k)} \right)^{-1} \Hc_{\pi(j)}^{\dag} \right]^{T} \hspace{-.1in}.
\end{eqnarray*}
To compute the gradient of $F(\Q)$ with respect to $\Q_{\pi(j)}$, we notice that only the first $j$
terms in $F(\Q)$ involve $\Q_{\pi(j)}$. From the definition $\nabla_{z} f(z) = 2(\partial
f(z)/\partial z)^{*}$ \cite{Haykin96:Adpt_Fltr}, we have
\begin{eqnarray} \label{eqn_gradient}
&& \hspace{-.25in} \bar{\G}_{\pi(j)} = 2\rho_{\pi(j)}\Hc_{\pi(j)} \Bigg[ \sum_{i=1}^{j}\left(
u_{\pi(i)} - u_{\pi(i-1)} \right) \times \nonumber\\
&& \hspace{-.2in} \left. \left( \I + \sum_{k=i}^{K} \rho_{\pi(k)} \Hc_{\pi(k)}^{\dag} \Q_{\pi(k)}
\Hc_{\pi(k)} \right)^{-1} \right] \Hc_{\pi(j)}^{\dag}.
\end{eqnarray}
\begin{rem}
It is important to point out that we can exploit the special structure in (\ref{eqn_gradient}) to
significantly reduce the computation complexity in the implementation of the algorithm. Note that
the most difficult part in computing $\bar{\G}_{\pi(j)}$ is the summation of the terms in the form
of $\Hc_{\pi(k)}^{\dag} \Q_{\pi(k)}\Hc_{\pi(k)}$. Without careful consideration, one may end up
computing such additions $j(2K+1-j)/2$ times for $\bar{\G}_{\pi(j)}$. However, notice that when $j$
varies, most of the terms in the summation are still the same. Thus, we can maintain a running sum
for $\I + \sum_{k=i}^{K} \rho_{\pi(k)} \Hc_{\pi(k)}^{\dag} \Q_{\pi(k)} \Hc_{\pi(k)}$, start out
from $j=K$, and reduce $j$ by one sequentially. As a result, only one new term is added to the
running sum in each iteration, which means we only need to do the addition once in each iteration.
\end{rem}

The conjugate gradient direction in the $m^{th}$ iteration can be computed as $\G_{\pi(j)}^{(m)} =
\bar{\G}_{\pi(i)}^{(m)} + \kappa_{m} \G_{\pi(i)}^{(m-1)}$. We adopt the Fletcher and Reeves' choice
of deflection \cite{Bazaraa_Sherali_Shetty_93:NLP}, which can be computed as
\begin{equation} \label{eqn_conjuate}
\kappa_{m} = \frac{\| \bar{\G}_{\pi(j)}^{(m)} \|^{2}}{\| \bar{\G}_{\pi(j)}^{(m-1)} \|^{2}}.
\end{equation}
The purpose of deflecting the gradient using (\ref{eqn_conjuate}) is to find $\G_{\pi(j)}^{(m)}$,
which is the Hessian-conjugate of $\G_{\pi(j)}^{(m-1)}$. By doing so, we can eliminate the
``zigzagging'' phenomenon encountered in the conventional gradient projection method, and achieve
the superlinear convergence rate \cite{Bazaraa_Sherali_Shetty_93:NLP} without actually storing a
large Hessian approximation matrix as in quasi-Newton methods.

\subsection{Projection onto $\Pwrcon$} Noting from (\ref{eqn_gradient}) that $\G_{\pi(j)}$ is
Hermitian, we have that $\Q_{\pi(j)}^{'(k)} = \Q_{\pi(j)}^{(k)}+s_{k} \G_{\pi(j)}^{(k)}$ is
Hermitian as well. Then, the projection problem becomes how to simultaneously project $|\Out{n}|$
Hermitian matrices onto the set
\begin{equation*}
\Pwrcon \triangleq \left\{ \Q_{l} \left|
\begin{array}{l}
\sum_{l} \tr\{ \Q_{l} \} \leq P_{\max}^{(n)}, \\
\Q_{l} \succeq 0, \,\, l \in \Out{n}
\end{array}
\right. \right\}.
\end{equation*}
This problem belongs to the class of ``matrix nearness problems'' \cite{Boyd05:Mtrx_Nrns,
Malick05:Mtrx_Nrns}, which are not easy to solve in general. However, by exploiting the special
structure in $\Theta_{\link}^{(n)}(\u)$, we are able to design a polynomial-time algorithm.

We construct a block diagonal matrix $\D = \diag{\Q_{\pi(1)} \ldots \Q_{\pi(K)}} \in
\mathbb{C}^{(K\cdot n_{r})\times(K\cdot n_{r})}$. It is easy to recognize that $\Q_{\pi(j)} \in
\Pwrcon$, $j=1,\ldots,K$, only if $\tr(\D) = \sum_{j=1}^{K} \tr \left(\Q_{\pi(j)}\right) \leq
P_{\max}^{(n)}$ and $\D \succeq 0$. We use Frobenius norm, denoted by $\|\cdot\|_{F}$, as the
matrix distance criterion. The distance between two matrices $\A$ and $\B$ is defined as $\| \A -
\B \|_{F} = \left( \tr\left[ (\A-\B)^{\dag} (\A-\B) \right] \right)^{\frac{1}{2}}$. Thus, given a
block diagonal matrix $\D$, we wish to find a matrix $\tilde{\D} \in \Pwrcon$ such that
$\tilde{\D}$ minimizes $\| \tilde{\D} - \D \|_{F}$. For more convenient algebraic manipulations, we
instead study the following equivalent optimization problem:
\begin{equation} \label{eqn_proj_primal_equiv}
\begin{array}{rl}
\mbox{Minimize} & \frac{1}{2} \| \tilde{\D} - \D \|_{F}^{2} \\
\mbox{subject to} & \tr (\tilde{\D}) \leq P_{\max}^{(n)}, \,\, \tilde{\D} \succeq 0. \\
\end{array}
\end{equation}
In (\ref{eqn_proj_primal_equiv}), the objective function is convex in $\tilde{\D}$, the constraint
$\tilde{\D} \succeq 0$ represents the convex cone of positive semidefinite matrices, and the
constraint $\tr (\tilde{\D}) \leq P_{\max}^{(n)}$ is a linear constraint. Thus, the problem is a
convex minimization problem and we can exactly solve this problem by solving its Lagrangian dual
problem. Associating Hermitian matrix $\mOmega$ to the constraint $\tilde{\D} \succeq 0$ and $\mu$
to the constraint $\tr (\tilde{\D}) \leq P_{\max}^{(n)}$, we can write the Lagrangian as
$g(\mOmega, \mu) = \min_{\tilde{\D}} \{ (1/2) \| \tilde{\D} - \D \|_{F}^{2} - \tr(\mOmega^{\dag}
\tilde{\D}) + \mu (\tr(\tilde{\D})- P_{\max}^{(n)} ) \}$. Since $g(\mOmega, \mu)$ is an
unconstrained convex quadratic minimization problem, we can compute the minimizer of the Lagrangian
by simply setting its first derivative (with respect to $\tilde{\D}$) to zero, i.e., $(\tilde{\D} -
\D) - \mOmega^{\dag} + \mu \I = 0$. Noting that $\mOmega^{\dag} = \mOmega$, we have $\tilde{\D} =
\D - \mu \I + \mOmega$. Substituting $\tilde{\D}$ back into the Lagrangian, we have
\begin{equation*}
g(\mOmega,\mu) = -\frac{1}{2} \norm{ \D - \mu\I + \mOmega }_{F}^{2} - \mu P_{\max}^{(n)} +
\frac{1}{2} \| \D \|^{2}.
\end{equation*}
Therefore, the Lagrangian dual problem can be written as
\begin{equation}\label{eqn_prj_dual}
\begin{array}{rl}
\mbox{Maximize} & -\frac{1}{2} \norm{ \D - \mu\I + \mOmega }_{F}^{2} - \mu P_{\max}^{(n)} + \frac{1}{2} \| \D \|^{2} \\
\mbox{subject to} & \mOmega \succeq 0, \mu \geq 0.
\end{array}
\end{equation}
After solving (\ref{eqn_prj_dual}), we can have the optimal solution to
(\ref{eqn_proj_primal_equiv}) as
\begin{equation*}
\tilde{\D}^{*} = \D - \mu^{*} \I + \mOmega^{*},
\end{equation*}
where $\mu^{*}$ and $\mOmega^{*}$ are the optimal dual solutions to Lagrangian dual problem in
(\ref{eqn_prj_dual}). We now consider the term $\D - \mu\I + \mOmega$, which is the only term
involving $\mOmega$ in the dual objective function. From Moreau Decomposition
\cite{Hiriart-Urruty_Lemarechal01:Cnvx_Anl}, we immediately have
\begin{equation*}
\min_{\mOmega} \norm{\D - \mu\I + \mOmega}_{F} = \left(\D - \mu\I \right)_{+},
\end{equation*}
where the operation $(\A)_{+}$ means performing eigenvalue decomposition on matrix $\A$, keeping
the eigenvector matrix unchanged, setting all non-positive eigenvalues to zero, and then
multiplying back. Thus, the matrix variable $\mOmega$ in the Lagrangian dual problem can be removed
and the Lagrangian dual problem can be rewritten as
\begin{equation}\label{eqn_prj_dual_sim}
\begin{array}{rl}
\!\!\!\!\!\!\! \mbox{Maximize} & \!\!\! \psi(\mu) \triangleq -\frac{1}{2} \norm{ \left(\D - \mu\I
\right)_{+} }_{F}^{2} - \mu P_{\max}^{(n)} \\
\!\!\!\!\!\!\! \mbox{subject to} & \!\!\! \mu \geq 0.
\end{array}
\end{equation}
Suppose that after performing eigenvalue decomposition on $\D$, we have $\D = \U \mLambda
\U^{\dag}$, where $\mLambda$ is the diagonal matrix formed by the eigenvalues of $\D$, $\U$ is the
unitary matrix formed by the corresponding eigenvectors. Since $\U$ is unitary, we have
\begin{equation*}
\left(\D - \mu\I \right)_{+} = \U \left(\mLambda - \mu \I \right)_{+} \U^{\dag}.
\end{equation*}
It then follows that
\begin{equation*}
\norm{ \left(\D - \mu\I \right)_{+} }_{F}^{2} = \norm{\left(\mLambda - \mu \I \right)_{+}}_{F}^{2}.
\end{equation*}
We denote the eigenvalues in $\mLambda$ by $\lambda_{i}$, $i=1,2,\ldots,K\cdot n_{r}$. Suppose that
we sort them in non-increasing order such that $\mLambda = \diag{\lambda_{1} \,\, \lambda_{2}
\ldots \,\, \lambda_{K\cdot n_{r}}}$, where $\lambda_{1} \geq \ldots \geq \lambda_{K\cdot n_{r}}$.
It then follows that
\begin{equation*}
\norm{\left(\mLambda - \mu \I \right)_{+}}_{F}^{2} = \sum_{j=1}^{K \cdot n_{r}} \left( \max
\left\{0, \lambda_{j} - \mu \right\} \right)^{2}.
\end{equation*}
So, we can rewrite $\psi(\mu)$ as
\begin{equation} \label{eqn_psi}
\psi(\mu) = - \frac{1}{2} \sum_{j=1}^{K \cdot n_{r}} \left( \max \left\{0, \lambda_{j} - \mu
\right\} \right)^{2} - \mu P_{\max}^{(n)}.
\end{equation}
It is evident from (\ref{eqn_psi}) that $\psi(\mu)$ is continuous and (piece-wise) concave in
$\mu$. Due to this special structure, we can search the optimal value of $\mu$ as follows. Let
$\hat{I}$ index the pieces of $\psi(\mu)$, $\hat{I}=0,1,\ldots,K\cdot n_{r}$. Initially we set
$\hat{I}=0$ and increase $\hat{I}$ subsequently. Also, we introduce $\lambda_{0} = \infty$ and
$\lambda_{K\cdot n_{r}+1} = -\infty$. We let the endpoint objective value
$\psi_{\hat{I}}\left(\lambda_{0} \right) = 0$, $\phi^{*} = \psi_{\hat{I}} \left( \lambda_{0}
\right)$, and $\mu^{*} = \lambda_{0}$. If $\hat{I} > K \cdot n_{r}$, the search stops. For a
particular index $\hat{I}$, by setting
\begin{equation*}
\frac{\partial}{\partial \mu} \psi_{\hat{I}}(\nu) \triangleq \frac{\partial}{\partial \mu} \left(
-\frac{1}{2} \sum_{i=1}^{\hat{I}}\left(\lambda_{i}-\mu \right)^{2} - \mu P_{\max}^{(n)} \right) =
0,
\end{equation*}
we have
\begin{equation*}
\mu_{\hat{I}}^{*} = \frac{\sum_{i=1}^{\hat{I}} \lambda_{i} - P_{\max}^{(n)}}{\hat{I}}.
\end{equation*}
Now we consider the following two cases: 
\begin{enumerate}
\item If $\mu_{\hat{I}}^{*} \in \left[\lambda_{\hat{I}+1}, \lambda_{\hat{I}}\right] \cap
\mathbb{R}_{+}$, where $\mathbb{R}_{+}$ denotes the set of non-negative real numbers, then
$\mu_{\hat{I}}^{*}$ is the optimal solution because $\psi(\mu)$ is concave in $\mu$. Thus, the
point having zero-value first derivative, if exists, must be the unique global maximum solution.
Hence, we can let $\mu^{*}=\mu_{\hat{I}}^{*}$ and the search is done.

\item If $\mu_{\hat{I}}^{*} \notin \left[\lambda_{\hat{I}+1},\lambda_{\hat{I}}\right] \cap \mathbb{R}_{+}$,
we must have that the local maximum in the interval $\left[\lambda_{\hat{I}+1},
\lambda_{\hat{I}}\right] \cap \mathbb{R}_{+}$ is achieved at one of the two endpoints. Note that
the objective value $\psi_{\hat{I}}\left( \lambda_{\hat{I}} \right)$ has been computed in the
previous iteration because from the continuity of the objective function, we have
$\psi_{\hat{I}}\left( \lambda_{\hat{I}} \right) = \psi_{\hat{I}-1}\left( \lambda_{\hat{I}}
\right)$. Thus, we only need to compute the other endpoint objective value
$\psi_{\hat{I}}\left(\lambda_{\hat{I}+1}\right)$. If
$\psi_{\hat{I}}\left(\lambda_{\hat{I}+1}\right) < \psi_{\hat{I}}\left(\lambda_{\hat{I}}\right) =
\phi^{*}$, then we know $\mu^{*}$ is the optimal solution; else let $\mu^{*} =
\lambda_{\hat{I}+1}$, $\phi^{*} = \psi_{\hat{I}}\left( \lambda_{\hat{I}+1} \right)$, $\hat{I} =
\hat{I} + 1$ and continue.
\end{enumerate}
Since there are $K \cdot n_{r}+1$ intervals in total, the search process takes at
most $K \cdot n_{r}+1$ steps to find the optimal solution $\mu^{*}$. Hence, this search is of
polynomial-time complexity $O(n_{r}K)$. After finding $\mu^{*}$, we can compute $\tilde{\D}^{*}$ as
\begin{equation}
\tilde{\D}^{*} = \left( \D - \mu^{*} \I \right)_{+} = \U \left( \mLambda - \mu^{*} \I \right)_{+}
\U^{\dag}.
\end{equation}
The projection of $\D$ onto $\Pwrcon$ is summarized in Algorithm~\ref{alg_rlt_prj}.
\begin{algorithm}
\caption{Projection onto $\Pwrcon$} \label{alg_rlt_prj}
\begin{algorithmic}
{\footnotesize
\STATE {\bf Initiation:} \\
\STATE \quad 1. Construct a block diagonal matrix $\D$. Perform eigenvalue decompo- \\
\STATE \quad \quad sition $\D = \U \mathbf{\Lambda} \U^{\dag}$, sort the eigenvalues in non-increasing order. \\
\STATE \quad 2. Introduce $\lambda_{0} = \infty$ and $\lambda_{K \cdot n_{t}+1} = -\infty$. Let
$\hat{I}=0$. Let the \\
\STATE \quad \quad endpoint objective value $\psi_{\hat{I}}\left(\lambda_{0}\right) = 0$, $\phi^{*}
= \psi_{\hat{I}}\left(\lambda_{0}\right)$, and $\mu^{*} = \lambda_{0}$.
\STATE {\bf Main Loop:} \\
\STATE \quad 1. If $\hat{I} > K \cdot n_{r}$, go to the final step; else let
$\mu_{\hat{I}}^{*} = (\sum_{j=1}^{\hat{I}} \lambda_{j} - P )/\hat{I}$. \\
\STATE \quad 2. If {$\mu_{\hat{I}}^{*} \in [\lambda_{\hat{I}+1}, \lambda_{\hat{I}} ] \cap
\mathbb{R}_{+}$}, then let $\mu^{*} = \mu_{\hat{I}}^{*}$ and go to the final step. \\
\STATE \quad 3. Compute $\psi_{\hat{I}}(\lambda_{\hat{I}+1})$. If $\psi_{\hat{I}}(\lambda_{\hat{I}+1}) < \phi^{*}$, then go to the final step; \\
\STATE \quad \quad else let $\mu^{*} = \lambda_{\hat{I}+1}$, $\phi^{*} = \psi_{\hat{I}}(\lambda_{\hat{I}+1})$, $\hat{I} = \hat{I} + 1$ and continue. \\
\STATE {\bf Final Step:} Compute $\tilde{\D}$ as $\tilde{\D} = \U \left( \mLambda - \mu^{*} \I
\right)_{+} \U^{\dag}$.
 }
\end{algorithmic}
\end{algorithm}

\subsection{Solving the Master Dual Problem}

\subsubsection{Cutting-Plane Method for Solving $\Theta(\u)$}
The attractive feature of the cutting-plane method
is its robustness, speed of convergence, and its simplicity in recovering primal feasible optimal
solutions. The primal optimal feasible solution can be exactly computed by averaging all the primal
solutions (may or may not be primal feasible) using the dual variables as weights
\cite{Bazaraa_Sherali_Shetty_93:NLP}. Letting $z=\Theta(\u)$, the dual problem is equivalent to
\begin{equation} \label{eqn_cutting_plane_1}
\begin{array}{rl}
\hspace{-.15in} \mbox{Minimize} & z \\
\hspace{-.15in} \mbox{subject to} & z \geq \sum_{f} \ln \left( s_{f} \right) + \sum_{l} u_{l}
\left(R_{l}(\Q) - \langle \1, \T^{T}\e_{l} \rangle \right)\\
\hspace{-.15in} & \u \geq 0,
\end{array}
\end{equation}
where $(\S, \T, \Q) \in \Psi $. Although (\ref{eqn_cutting_plane_1}) is a linear program with
infinite constraints not known explicitly, we can consider the following {\em approximating}
problem:
\begin{equation} \label{eqn_cutting_plane_approx}
\begin{array}{rl}
\hspace{-.15in} \mbox{Minimize} & z \\
\hspace{-.15in} \mbox{subject to} & z \geq \sum_{f} \ln ( s_{f}^{(j)} ) + \sum_{l} u_{l} \left(R_{l}(\Q^{(j)}) - \right. \\
\hspace{-.15in} & \quad \quad \left. \langle \1, \T^{(j)T} \e_{l} \rangle \right)\!\!\!\\
\hspace{-.15in} & \u \geq 0,
\end{array}
\end{equation}
where the points $(\S^{(j)}, \T^{(j)}, \Q^{(j)}) \in \Psi$, $j=1,\ldots,k-1$. The problem in
(\ref{eqn_cutting_plane_approx}) is a linear program with a finite number of constraints and can be
solved efficiently. Let $(z^{(k)}, \u^{(k)})$ be an optimal solution to the approximating problem,
which we refer to as the {\em master program}. If the solution is feasible to
(\ref{eqn_cutting_plane_1}), then it is an optimal solution to the Lagrangian dual problem. To
check the feasibility, we consider the following {\em subproblem}:
\begin{equation} \label{eqn_cutting_plane_subprob}
\begin{array}{rl}
\mbox{Maximize} & \sum_{f} \ln \left( s_{f} \right) + \sum_{l} u_{l}^{(k)}
\left(R_{l}(\Q) - \langle \1, \T^{T} \e_{l} \rangle \right) \\
\mbox{subject to} & (\S, \T, \Q) \in \Psi
\end{array}
\end{equation}
Suppose that $(\S^{(k)}, \T^{(k)}, \Q^{(k)})$ is an optimal solution to the subproblem
(\ref{eqn_cutting_plane_subprob}) and $\Theta^{*}(\u^{(k)})$ is the corresponding optimal objective
value. If $z_{k} \geq \Theta^{*}(\u^{(k)})$, then $\u^{(k)}$ is an optimal solution to the
Lagrangian dual problem. Otherwise, for $\u = \u^{(k)}$, the inequality constraint in
(\ref{eqn_cutting_plane_1}) is not satisfied for $(\S^{(j)}, \T^{(j)}, \Q^{(j)})$. Thus, we can add
the constraint
\begin{equation} \label{eqn_cutting_plane_cut}
z \geq \sum_{f} \ln \left( s_{f}^{(k)} \right) + \sum_{l} u_{l} \left(R_{l}(\Q^{(k)}) - \langle \1,
\T^{(k)T} \e_{l} \rangle \right)
\end{equation}
to (\ref{eqn_cutting_plane_approx}), and re-solve the master linear program. Obviously, $(z^{(k)},
\u^{(k)})$ violates (\ref{eqn_cutting_plane_cut}) and will be cut off by
(\ref{eqn_cutting_plane_cut}). The cutting plane algorithm is summarized in
Algorithm~\ref{alg_cutting_plane}.
\begin{algorithm}
\caption{Cutting Plane Algorithm for Solving \dual} \label{alg_cutting_plane}
\begin{algorithmic}
{\footnotesize \STATE Initialization: \\
\STATE \quad Find a point $(\S^{(0)}, \T^{(0)}, \Q^{(0)}) \in \Psi$. Let $k=1$.\\
\STATE Main Loop: \\
\STATE \quad 1. Solve the master program in (\ref{eqn_cutting_plane_approx}). Let $(z^{(k)},
\u^{(k)})$ be an optimal \\
\STATE \quad \quad solution. \\
\STATE \quad 2. Solve the subproblem in (\ref{eqn_cutting_plane_subprob}). Let $(\S^{(k)},
\T^{(k)}, \Q^{(k)})$ be an optimal \\
\STATE \quad \quad point, and let $\Theta^{*}(\u^{(k)})$ be the corresponding optimal objective value.\\
\STATE \quad 3. If $z^{(k)} \geq \Theta(\u^{(k)})$, then stop with $\u^{(k)}$ as the optimal dual
solution. \\
\STATE \quad \quad Otherwise, add the constraint (\ref{eqn_cutting_plane_cut}) to the master
program, replace $k$ \vspace{-.05in} \\
\STATE \quad \quad by $k+1$, and go to step 1. }
\end{algorithmic}
\end{algorithm}


\subsubsection{Subgradient Algorithm for Solving $\Theta(\u)$}
Since the Lagrangian dual objective function is piece-wise differentiable, subgradient method can
also be applied. For $\Theta(\u)$, starting with an initial $\u^{(1)}$ and after evaluating
subproblems $\Theta_{\net}(\u)$ and $\Theta_{\link}$ for $\u^{(k)}$ in the $k^{th}$ iteration, we
update the dual variables by $\u^{(k+1)} = \left[ \u^{k} - \lambda_{(k)} \mathbf{d}^{(k)}
\right]_{+}$, where the operator $[\cdot]_{+}$ projects a vector on to the nonnegative orthant, and
$\lambda_{k}$ denotes a positive scalar step size. $\mathbf{d}^{(k)}$ is a subgradient of the
Lagrangian at point $\u^{(k)}$. It is proved in \cite{Bazaraa_Sherali_Shetty_93:NLP} that the
subgradient algorithm converges if the step size $\lambda_{k}$ satisfies $\lambda_{k} \rightarrow
0$ as $k \rightarrow \infty$ and $\sum_{k=0}^{\infty} \lambda_{k} = \infty$. A simple and useful
step size selection strategy is the divergent harmonic series $\sum_{k=1}^{\infty} \beta
\frac{1}{k} = \infty$, where $\beta$ is a constant. The subgradient for the Lagrangian dual problem
can be computed as
\begin{equation} \label{eqn_subgrad}
R_{l}(\Q^{*}(\u)) - \langle \1, \T^{*}(\u)^{T}\e_{l} \rangle, \quad l=1,2,\ldots,L.
\end{equation}
Specifically, the subgradient method has the following properties which make it possible to be
implemented in a {\em distributed} fashion:
\begin{enumerate}
\item Subgradient computation only requires local traffic information $\langle \1,
\T^{T}\e_{l}\rangle$ and the available link capacity information $R_{l}(\Q)$ at each link $l$. As a
result, it can be computed locally.

\item The choice of step size $\lambda_{k} = \beta \frac{1}{k}$
depends only upon the iteration index $k$, and does not require any other global knowledge. In
conjunction with the first property, the dual variable, in the iterative form of $u_{l}^{(k+1)} =
u_{l}^{(k)} + \lambda_{k} (\partial \Theta(\u)/\partial u_{l})$, can also be computed locally.

\item The objective functions $\Theta_{\link}$ can be decomposed on a node-by-node basis such that each node in
the network can perform the computation in parallel. Likewise, the network layer subproblem
$\Theta_{\net}$ can be decomposed on a source-by-source basis such that each source node can
perform the routing computation locally after receiving the dual variable information of each link
in the network.
\end{enumerate}

It is worth to point out that care must be taken when recovering the primal feasible optimal
solution in the subgradient method. Generally, the primal variables in the dual optimal solution
are not primal feasible unless the dual optimal solution happens to be the saddle point.
Fortunately, since CRPA-E is convex, its primal feasible optimal solution can be exactly computed
by solving a linear programming problem (see \cite{Bazaraa_Sherali_Shetty_93:NLP} for further
details). However, such a recovery approach cannot be implemented in a distributed fashion. In this
paper, we adopt a variant of Shor's rule to recovery primal optimal feasible solution. Due to space
limitation, we refer readers to \cite{Sherali_ORL96} for more details.

\section{Numerical Results} \label{sec:numerical}
\begin{figure} 
\centering
\includegraphics[width=2.5in]{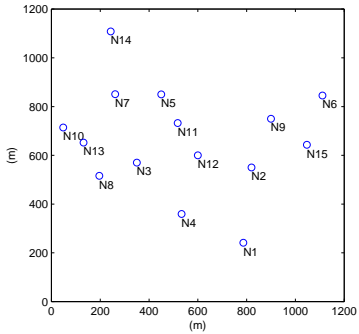}
\caption{A 15-node network example.} \label{fig_topo66}
\end{figure}

In this section, we present some numerical results through simulations to provide further insights
on solving CRPA. $N$ randomly-generated MIMO-enabled nodes are uniformly distributed in a square
region. Each node in the network is equipped with two antennas. The maximum transmit power for each
node is set to $P_{\max} = 10$dBm. Each node in the network is assigned a unit bandwidth. We
illustrate a 15-node network example, as shown in Fig.~\ref{fig_topo66}, to show the convergence
process of the cutting-plane and the subgradient methods for solving
$\mathbf{D}^{\mathrm{CRPA-E}}$. In this example, there are three flows transmitting across the
network: N14 to N1, N6 to N10, and N5 to N4, respectively.

\subsection{Cutting-Plane Method}
\begin{figure} 
\centering
\includegraphics[width=2.8in]{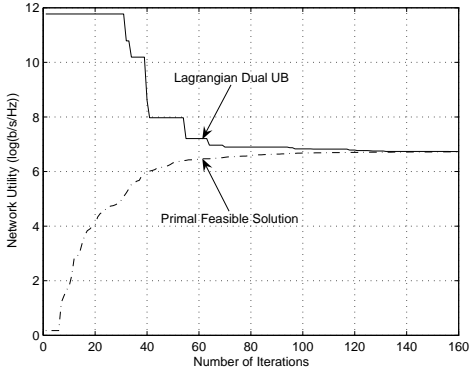}
\caption{Convergence behavior of the cutting-plane method} \label{fig_cp66}
\end{figure}

For the 15-node example in Fig.~\ref{fig_topo66}, the convergence process for the cutting-plane
method is illustrated in Fig.~\ref{fig_cp66}.
The optimal objective value for this 15-node example is 6.72. The optimal flows for sessions N14 to
N1, N6 to N10, and N5 to N4 are $9.17$ bps/Hz, $9.30$ bps/Hz, and $9.93$ bps/Hz, respectively. It
can be observed that the cutting-plane algorithm is very efficient: It converges with approximately
160 cuts. As expected, the duality gap is zero because the convexity of the transformed equivalent
problem based on dual MIMO-MAC.

\subsection{Subgradient Method}
\begin{figure} 
\centering
\includegraphics[width=2.8in]{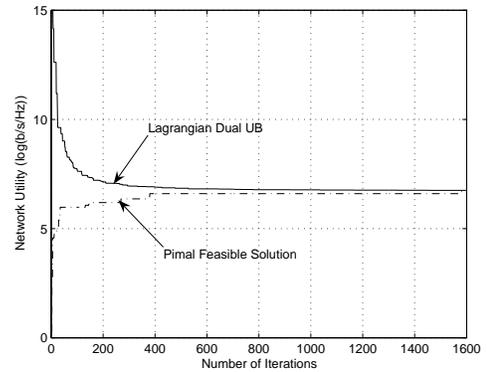}
\caption{Convergence behavior of the subgradient method} \label{fig_sg66}
\end{figure}

For the 15-node example in Fig.~\ref{fig_topo66}, the convergence process for the subgradient
method is illustrated in Fig.~\ref{fig_sg66}.
The step size selection is $\lambda_{k} = 0.1/k$. The subgradient method also achieves the same
optimal solution and objective value when it converges. However, it is seen that the subgradient
algorithm takes approximately 1600 iterations to converge, which is much slower than the
cutting-plane method. This is partially due to the heuristic nature in step size selection (cannot
be too large or too small at each step). It is also partially due to the cumbersomeness in
recovering the primal feasible solution in the subgradient method. In this example, the dual upper
bound takes approximately 1050 iterations to reach near the optimal. However, the near-optimal
primal feasible solution cannot be identified until after 1500 iterations.

\subsection{Comparison between BC and TDM}
\begin{figure} 
\centering
\includegraphics[width=2.8in]{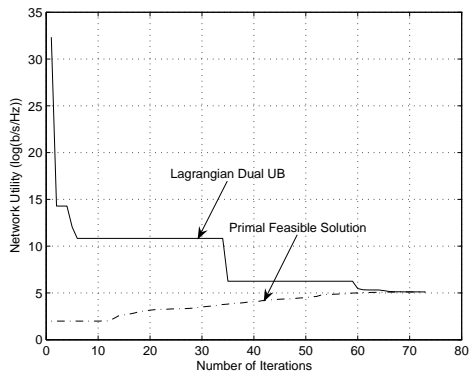}
\caption{The convergence behavior in TDM case} \label{fig_cp66_tdm}
\end{figure}

We now study how much performance gain we can get by using Gaussian vector broadcast channel
technique as opposed to the conventional time-division (TDM) scheme. The cross-layer optimization
problem of MIMO-based mesh networks over TDM scheme is also a convex problem. Thus, the basic
Lagrangian dual decomposition framework and gradient projection technique for the link layer
subproblem are still applicable. The only difference is in the gradient computation, which is
simpler in TDM case. For the same 15-node network with TDM, we plot the convergence process of the
cutting-plane algorithm in Fig.~\ref{fig_cp66_tdm}.
In TDM case, the optimal objective value is 5.01. For this example, we have $34.4\%$ improvement by
using DPC.

\section{Related Work} \label{sec:related}
Despite significant research progress in using MIMO for single-user communications, research on
multi-user multi-hop MIMO networks is still in its inception stage. There are many open problems,
and many areas are still poorly understood \cite{Goldsmith03:MIMO_Limit}. Currently, the relatively
well-studied research area of multi-user MIMO systems are cellular systems, which are single-hop
and infrastructure-based. For multi-hop MIMO-based mesh networks, research results remain limited.
In \cite{Hu04:MIMO_Hop_Dist}, Hu and Zhang studied the problem of joint medium access control and
routing, with a consideration of optimal hop distance to minimize end-to-end delay. In
\cite{Sundaresan05:MIMO_Routing}, Sundaresan and Sivakumar used simulations to study various
characteristics and tradeoffs (multiplexing gain vs. diversity gain) of MIMO links that can be
leveraged by routing layer protocols in rich multipath environments to improve performance. In
\cite{Lee06:Rate_Reliability}, Lee et al. proposed a distributed algorithm for MIMO-based multi-hop
ad hoc networks, in which diversity and multiplexing gains of each link are controlled to achieve
the optimal rate-reliability tradeoff. The optimization problem assumes fixed SINRs and fixed
routes between source and destination nodes. However, in these works, there is no explicit
consideration of per-antenna power allocation and their impact on upper layers. Moreover, DPC in
cross-layer design has never been studied either.

\section{Conclusions} \label{sec:conclusion}
In this paper, we investigated the cross-layer optimization of DPC per-antenna power allocation and
multi-hop multi-path routing for MIMO-based wireless mesh networks. Our contributions are
three-fold. First, this paper is the first work that studies the impacts of applying dirty paper
coding, which is the optimal transmission scheme for MIMO broadcast channels (MIMO-BC), to the
cross-layer design for MIMO-based wireless mesh networks. We showed that the network performance
has dramatic improvements compared to that of the conventional time-division/frequency division
schemes. Second, we solved the challenging non-connvex cross-layer optimization problem by
exploiting the channel duality between MIMO-MAC and MIMO-BC, and we showed that transformed problem
under dual MIMO-MAC is convex. We simplified the maximum weighted sum rate problem, thus paving the
way for solving the link layer subproblem in the Lagrangian dual decomposition. Last, for the
transformed problem, we develop an efficient solution procedure that integrates Lagrangian dual
decomposition, conjugate gradient projection based on matrix differential calculus, cutting-plane,
and subgradient methods. Our results substantiate the importance of cross-layer optimization for
MIMO-based wireless mesh networks with Gaussian vector broadcast channels.

\end{document}